\documentclass{SCIS2025}

\begin{document}
\ArticleType{RESEARCH PAPER}
\Year{2025}
\Month{January}
\Vol{68}
\No{1}
\DOI{}
\ArtNo{}
\ReceiveDate{13-Dec-2024}
\ReviseDate{11-Apr-2025}
\AcceptDate{19-May-2025}
\OnlineDate{}
\AuthorMark{Luo Y}
\AuthorCitation{Luo Y, Shi L H, Li Y H, et al}

\title{From Intention to Implementation: Automating Biomedical Research via LLMs}{From Intention to Implementation: Automating Biomedical Research via LLMs}

\author[1]{Yi Luo}{}
\author[1]{Linghang Shi}{}
\author[1]{Yihao Li}{}
\author[2]{Aobo Zhuang}{}
\author[3]{Yeyun Gong}{yegong@microsoft.com}
\author[4]{Ling Liu}{}
\author[1,5]{Chen Lin}{chenlin@xmu.edu.cn}


\address[1]{School of Informatics, National Institute for Data Science in Health and Medicine, Xiamen University, Xiamen {\rm 361101}, China}
\address[2]{School of Medicine, Xiamen University, Xiamen {\rm 361101}, China}
\address[3]{Microsoft Research Asia, Beijing {\rm 100080}, China}
\address[4]{College of Computing, Georgia Institute of Technology, Atlanta {\rm 30332}, USA}
\address[5]{Zhongguancun Academy, Beijing {\rm 100094}, China}
\newcommand{\system}{\textbf{BioResearcher}}

\abstract{Conventional biomedical research is increasingly labor-intensive due to the exponential growth of scientific literature and datasets. Artificial intelligence (AI), particularly Large Language Models (LLMs), has the potential to revolutionize this process by automating various steps. Still, significant challenges remain, including the need for multidisciplinary expertise, logicality of experimental design, and performance measurements.
This paper introduces BioResearcher, the first end-to-end automated system designed to streamline the entire biomedical research process involving dry lab experiments. BioResearcher employs a modular multi-agent architecture, integrating specialized agents for search, literature processing, experimental design, and programming. By decomposing complex tasks into logically related sub-tasks and utilizing a hierarchical learning approach, BioResearcher effectively addresses the challenges of multidisciplinary requirements and logical complexity. Furthermore, BioResearcher incorporates an LLM-based reviewer for in-process quality control and introduces novel evaluation metrics to assess the quality and automation of experimental protocols.
BioResearcher successfully achieves an average execution success rate of 63.07\% across eight previously unmet research objectives. The generated protocols, on average, outperform typical agent systems by 22.0\% on five quality metrics. The system demonstrates significant potential to reduce researchers' workloads and accelerate biomedical discoveries, paving the way for future innovations in automated research systems.}


\maketitle

\section{Introduction}\label{sec:intro}

Biomedical research is a fundamental driving force behind human development. By uncovering the underlying mechanisms of diseases~\cite{1}, biomedical research improves global health, extends life expectancy, and enhances life quality. It also fuels economic growth, scientific advancements, and societal well-being. 

Traditional biomedical research relies heavily on labor-intensive processes like manual data collection, comprehensive literature reviews, complex experimental designs, and extensive data analysis. Although the conventional approach has facilitated notable breakthroughs in disease prevention~\cite{2, 3, 4}, diagnosis~\cite{5, 6, 7}, and treatment~\cite{8,9,10}, it struggles to keep pace with the data explosion. For example, PubMed now hosts over 37 million citations~\footnote{\url{https://pubmed.ncbi.nlm.nih.gov/?Db=pubmed}}, overwhelming researchers and making it challenging to stay updated with the latest findings. Moreover, traditional research often demands repetitive tasks or interdisciplinary skills, such as coding, hindering research efficiency.

Artificial intelligence (AI) is emerging as a valuable tool in biomedical research, enhancing specific steps in the research pipeline. For instance, novel machine learning models are designed to analyze data and make decisions, including predicting drug-target from compound-protein interactions~\cite{11, 12, 13}, detecting the presence of cancer from medical images~\cite{14, 15}, estimating patient outcomes from medical history~\cite{16, 17, 18}, and so on. The advancement of Large Language Models (LLMs) further supports AI's role in academic writing~\cite{19, 20} and literature summarizing~\cite{21}. However, biomedical research is still time-consuming. The potential for automating the entire biomedical research process remains largely unexplored~\cite{22}, which allows researchers to concentrate on innovation and strategic decision-making.

Despite the potential of AI, several critical challenges must be overcome to build a fully automated biomedical research assistant. 

\begin{figure}[h!]
  \centering
  \includegraphics[width=\textwidth]{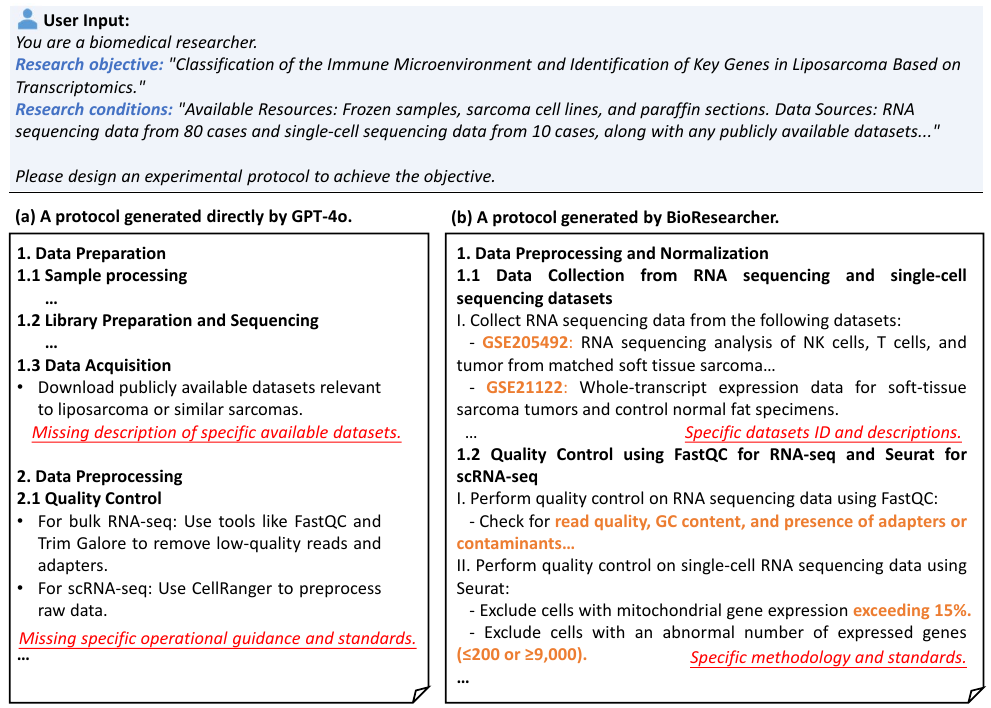}
  \caption{Example of experimental protocol generation. For the same user input, (a) illustrates an experimental protocol generated by a single LLM (GPT-4o), while (b) demonstrates a protocol generated by \system.}
  \label{fig:case}
\end{figure}

Firstly, biomedical research demands a \textbf{multidisciplinary skill set}, including a fundamental understanding of biology and medicine, comprehension of literature for available datasets and effective approaches, proficiency in programming languages to write code, knowledge of statistics to interpret results, etc. General-purpose LLMs like OpenAI's GPT-4o lack the domain-specific expertise needed for these tasks. For instance, as illustrated in Figure~\ref{fig:case}(a), when GPT-4o is asked to design a biomedical protocol for a specific research objective, the response is not executable due to missing details such as datasets and operational guidance.

Secondly, biomedical research is \textbf{logically complex}. On one hand, it requires a coherent understanding of literature with intricate logical structures, yet existing LLMs perform poorly on critical literature analysis. For example, irrelevant information in lengthy research papers causes catastrophic forgetting of important facts. On the other hand, biomedical research involves breaking down complex problems into logically related subtasks. For example, studying pyroptosis in dedifferentiated liposarcoma (DDL) involves multiple interconnected tasks like analyzing the expression variation and genetic changes of PRGs, performing immune infiltration analyses, identifying PRG-related clusters, characterizing the tumor microenvironment within these clusters, and developing a prognostic gene model based on these clusters~\cite{23}. Each task is logically interdependent, making planning and execution by LLMs challenging.

Lastly, it is crucial to \textbf{measure the performance} of the research assistant. From the quality control perspective, assessing the results of intermediate steps ensures that the research assistant provides reliable final outputs. From the evaluation perspective, detecting the errors of end-to-end responses helps to identify the weaknesses and strengths of different systems and sheds insight for future improvements. Due to the complexity of biomedical research tasks, manually conducting fine-grained evaluation is infeasible; for example, researchers are not patient with labeling errors in an experimental protocol line by line. Current automatic evaluations, such as ROUGE~\cite{24} and BLEU~\cite{25}, compute the overlap between the LLM's generation and the ground-truth answer, which is unavailable. Furthermore, they focus on textual quality while ignoring aspects such as technical completeness and correctness, which are more important in assessing the quality of experiment protocols. 

This paper introduces \system\footnote{Our code and prompts are available at \url{https://github.com/XMUDM/BioResearcher}.}, an intelligent research assistant designed to \textbf{automate the entire biomedical research process}. \system~can take any research objective and conditions the user provides, survey relevant literature, design an appropriate experiment protocol\footnote{Our effort primarily focuses on dry lab experiments, which rely on computational methods to conduct bioinformatics analysis. Future research endeavors may extend to wet lab experiments.}, write programs to implement the protocol and derive meaningful conclusions. \system~develops novel techniques to address the three challenges above. 

\system~employs a modular multi-agent architecture to integrate multidisciplinary skills. It comprises four modules: \texttt{Search}, \texttt{Literature Processing}, \texttt{Experimental Design}, and \texttt{Programming}, each containing multiple specialized agents. These agents specialize in distinct tasks, including literature and dataset search, filtering, reports generation from literature, reports analysis, experimental protocol design, dry lab experiment extraction, code writing and execution, and review. The specialization adapts LLMs to different task requirements, e.g., a search agent is more professional in retrieving relevant literature than a general-purpose LLM. The collaboration among multiple agents and modules increases the overall performance. As illustrated in Figure~\ref{fig:case}(b), the \texttt{Experimental Design} module generates a comprehensive experimental protocol detailing specific datasets, methodologies, and standards. This achievement is facilitated by the effective collection, processing, and analysis of relevant literature and datasets by the \texttt{Search} and \texttt{Literature Processing} modules. Compared with agent systems employing planning agents powered by LLMs to determine agent participation and intervention strategies throughout task execution~\cite{26, 27}, our system adopts a professionally designed and rigorously structured workflow framework that constrains agent generation processes through systematic procedural constraints, thereby ensuring enhanced stability and reproducibility. Furthermore, our framework diverges from existing human-curated multi-agent systems~\cite{28, 29, 30}. In computer science and other fields, current studies often prioritize novelty. However, due to the nature of our biomedical applications, we introduce the Literature Processing module to ensure the reliability and feasibility of our approach in the biomedical domain.

\system~adopts a hierarchical learning approach to decompose the complex logical structure. The \texttt{Literature Processing} module standardizes relevant papers into experimental reports to minimize unimportant information and provides analyses. The \texttt{Experimental Design} module then uses the Retrieval-Augmented Generation (RAG)~\cite{26} technique, aided by the analyses, to learn knowledge at different levels of granularity, including relevant headings, outlines, and experimental details, thereby facilitating the design of new protocols in a stepwise manner.

\system~introduces an LLM-based reviewer to provide feedback and refine itself for in-process quality control. This approach allows for the ongoing assessment of the generated content, ensuring it meets quality standards and aligns with research objectives. Moreover, we propose new evaluation metrics to assess the quality of the end-to-end performance, including five dimensions for protocol quality—completeness, level of detail, correctness, logical soundness, and structural soundness—and two metrics for experimental automation: execution success rate and error level.

Our \textbf{contributions} are summarized as follows:
\begin{itemize}
\item We present the first end-to-end automated system designed specifically for biomedical research. At its core is a multi-agent framework powered by LLMs, which decomposes complex research tasks into specialized subtasks. By enabling collaborative execution among domain-specific agents, the system enhances overall performance, significantly reduces manual effort, and improves efficiency.
\item We propose new evaluation metrics to assess the quality of the end-to-end performance, including five dimensions for protocol quality and two metrics for experimental automation.
\item Our system successfully achieves an average execution success rate of 63.07\% across eight previously unmet research objectives composed by senior researchers, with protocols outperforming typical agent systems by 22.0\% regarding the proposed quality metrics.
\item This study explores \system's potential to automate biomedical research, paving the way for future innovations and accelerating discoveries.
\end{itemize}

\section{Related Work}\label{sec:relatedwork}

Artificial intelligence (AI) has improved biomedical applications for processing different textual data~\cite{31, 32, 33}. Conventionally, small-scale language models~\cite{34} were used, but scaling laws indicate that increasing model parameters brings enhanced performance, leading to superior reasoning capability and higher answer accuracy. Therefore, we have seen large amounts of applications based on large language models (LLMs)~\cite{35, 36, 37, 38, 39}. 

\textbf{Biomedical LLMs}. 
Large language models (LLMs) are generally pre-trained on vast open-domain corpora but lack domain-specific knowledge. To enhance domain-specific performance, several techniques are used: (1) Fine-tuning optimizes the total or a part of parameters to improve the model performance on a small, specific dataset~\cite{35, 37, 38, 40}. (2) Reinforcement learning with human feedback (RLHF) or AI feedback (RLAIF) updates the model parameters to align LLM’s responses with responses from humans or a teacher-model via a reinforcement learning framework~\cite{36, 39, 41}. (3) Prompt engineering~\cite{42} involves giving instructions or examples to LLM to enforce rules or enhance reasoning~\cite{43, 44, 45}. These methods mainly focus on question-answering (QA) tasks, like answering medical consultations~\cite{36}, taking medical examinations~\cite{35} and summarizing clinical reports~\cite{33}. However, LLMs still struggle with complex tasks that require more profound understanding and reasoning\cite{22}. Compared with QA tasks, responses for scientific research tasks are much longer and more professional and involve intricate, long logical chains. Fine-tuning and RLHF are infeasible due to the lack of instruction data, and simple prompt engineering cannot empower LLM with deep logical reasoning for professional topics.

\textbf{LLM-based Agents for Research}. 
LLM-based agents offer significant advantages over the direct use of LLMs, such as actively acquiring information, interacting with environments, and stronger reasoning and planning~\cite{22}. LLMs can function as a \textit{single agent} by being assigned with specific roles through in-domain fine-tuning~\cite{29} or role-specific prompts~\cite{33, 46, 47, 48, 49, 50, 51}.
In contrast, a \textit{Multi-Agent System (MAS)} comprises multiple LLM-based agents, enabling task completion through various cooperative methods. One approach involves assigning distinct roles to agents, who then reach a consensus through negotiation, like discussing clinical diagnosis via several medical experts~\cite{52, 53, 54}. However, the consensus can be unreliable due to potential instability and hallucinations from LLMs. Alternatively, MAS can distribute complex tasks into sub-tasks among agents, such as dividing a scientific discovery process into creating ideas, experimentation, and writing~\cite{28}.
MAS is promising for tasks with complex logical chains and highly detailed requirements. After decomposing tasks and assigning them to agents, we can synthesize the results effectively. Furthermore, employing a professional and rigorous workflow framework helps constrain the agents’ generative processes, ensuring highly reliable outcomes.

\textbf{AI for Research}. 
AI for Research (AI4R)~\cite{55} has gained significant attention, particularly in automating scientific workflows. Existing studies are categorized into four automation levels~\cite{22}: Level 0 automation performs specific predefined tasks~\cite{56}; Level 1 automation designs simple experimental protocols with in-silico or lab tools~\cite{46, 57}; Level 2 automation develops rigorous experimental protocols and employs statistical methods for hypothesis evaluation~\cite{28, 58}. Level 3 agents, which remain undeveloped, are envisioned to discover new methods and employ diverse techniques to measure biological phenomena.

Our work belongs to Level 2 automation and differs from current AI4R systems. (1) The system is not limited to solving a specific biomedical task. For example, CRISPR-GPT~\cite{59} customizes the workflow only for gene editing experiments. (2) We automate the entire process without manual maintenance of a template database like Genesis~\cite{60}. (3) Most level 2 AI4R systems are designed for computer science (CS)~\cite{28, 29, 30, 61, 62}, where public datasets like OpenReview are available, facilitating training and evaluation. Such extensive review data is lacking in the biomedical domain. (4) Existing studies emphasize novelty while we focus on reliability and feasibility, necessitating fine-grained literature analysis. For example, AIScientist~\cite{28} improves existing solutions based on a given task and initial experimental code. Conversely, \system~designs a series of executable experiments for a new research subject.

\section{\system}\label{sec:framework}
\begin{figure}[h!]
    \centering
    \includegraphics[width=\textwidth]{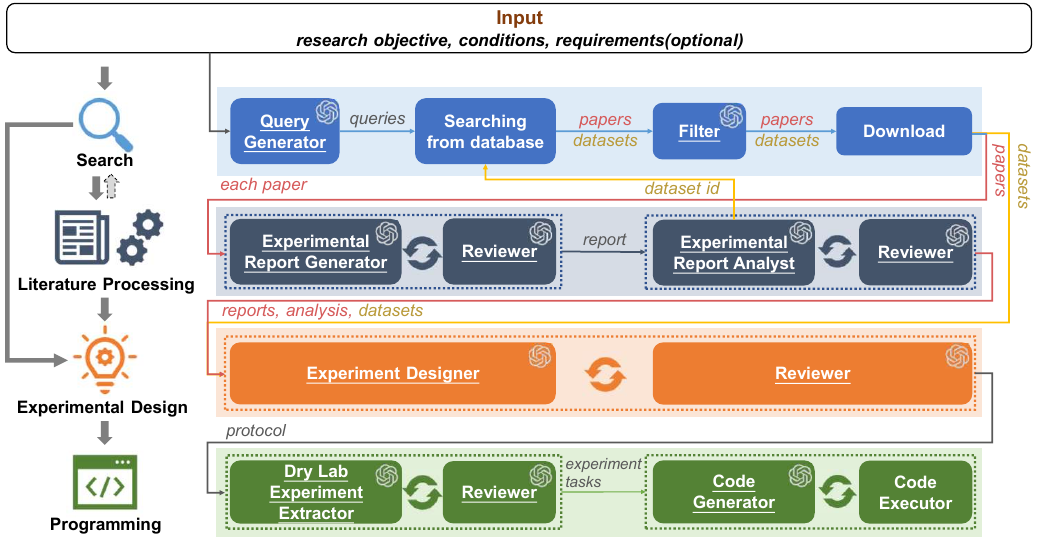}
    \caption{The flowchart of \system. On the left, the system is divided into four main modules: \texttt{Search}, \texttt{Literature Processing}, \texttt{Experimental Design}, and \texttt{Programming}. The right side illustrates the system's workflow: it initiates with the \texttt{Search} module that retrieves pertinent literature and datasets based on the user input. The \texttt{Literature Processing} module then converts the obtained literature into standardized experimental reports, analyzing them against the research objective, conditions, and requirements. It also works with the \texttt{Search} module to evaluate the applicability of the datasets mentioned in the reports. These processed reports, analyses, and suitable datasets are then sent to the \texttt{Experimental Design} module, which designs an experimental protocol aligned with the research objective. Finally, the \texttt{Programming} module extracts the dry lab experimental tasks from the protocol and generates executable code to perform these tasks. \underline{Components underlined} are agents based on LLMs.}
    \label{fig:framework}
\end{figure}
\subsection{Framework Overview} 
\system~is designed to automate the research process for biomedical studies. Users provide a research objective\footnote{Since our goal is to automate the entire research process, \system~currently only supports studies involving only dry lab
 experiments. Studies that require wet experiments are out of the scope because wet experiments generally require human operation, e.g., hands-on tasks in a laboratory setting, and can not be fully automated.}  and the conditions under which the experiments will be conducted. Users can also specify the research requirements, such as desired experimental steps or outcomes. All user input is in natural language. 

To emulate the workflow of human researchers, \system~consists of four primary modules: \texttt{Search, Literature Processing, Experimental Design, and Programming}. As shown in Figure~\ref{fig:framework}, the research process begins with the \texttt{Search} module, which comprehends the user input, generates appropriate queries, and searches for relevant research papers and datasets from online repositories. The retrieved literature is then filtered, downloaded, and forwarded to the \texttt{Literature Processing} module. Here, each research paper is standardized into an experimental report, and each report is analyzed in light of the user-specified objective, conditions, and requirements. This module also interacts with the \texttt{Search} module to identify the usability of the datasets mentioned in the reports. The processed reports, analyses, and datasets are then forwarded to the \texttt{Experimental Design} module, which constructs an experimental protocol. Finally, the \texttt{Programming} module extracts a sequence of dry lab experiment tasks from the protocol and generates accurate and executable codes for these tasks.

\subsection{Search}\label{sec:search}
\begin{figure}[h!]
    \centering
    \includegraphics[width=\textwidth]{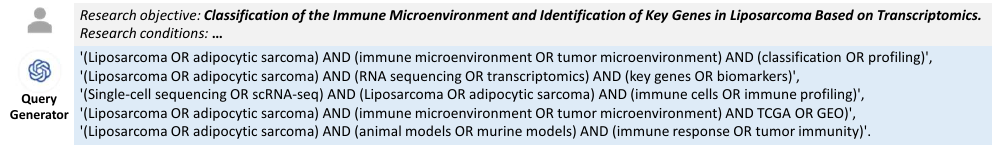}
    \caption{An example of query generation.}
    \label{fig:query_case}
\end{figure}

Scientific research initiates with a thorough review of related literature. Literature searching is especially important in \system~because, unlike previous studies that use literature surveys to identify research gaps and propose novel ideas, \system~ensures the process is built upon existing knowledge and preserves the reliability of the output. The \texttt{Search} module comprehensively explores pertinent literature and datasets throughout various research phases. Its internal procedure is as follows:

\textbf{(1) Query Generation:} 
Searching directly with the user input can yield imprecise results for two reasons. Firstly, the user describes the research objective and conditions in natural language, while most databases use Boolean query logic, making it difficult to retrieve results that completely match the lengthy user input. Secondly, the user intention is provided with key high-level concepts, while the literature may use a more detailed description or synonyms, making it challenging to obtain all relevant materials. Thus, query rewriting is crucial~\cite{63}. However, manually crafting effective queries costs a lot of expertise, labor, and time~\cite{64}. To address this, the \texttt{Search} module in \system~employs an LLM-based query generator agent to create Boolean queries based on the user input. As displayed in Figure~\ref{fig:query_case}, the query generator extracts keywords to improve the retrieval accuracy. It also performs synonym expansion (e.g., expanding "Single-cell sequencing" with "scRNA-seq") to improve the retrieval recall. These structured queries allow the module to interpret the research objective effectively and focus on the most relevant materials.

\textbf{(2) Retrieval:} The module interfaces with databases through their APIs, enabling the retrieval of relevant literature and datasets from established repositories such as PubMed Central (PMC), PubMed, and GEO. Additional databases can be integrated as needed to expand the system's capabilities.

\textbf{(3) Filtration:} To retain only the most relevant and useful materials for further stages, our system employs an LLM-based filter agent to filter the returned research articles and datasets. For the \textit{research articles}, we define a set of criteria detailed in Appendix D (Table D1). The filter examines the titles and abstracts of each article to determine their potential contribution to the research objectives. Each article receives a helpfulness score ranging from one to five. Articles scoring above four are downloaded and forwarded to the \texttt{Literature Processing} module. For the \textit{datasets}, the filter assesses its usefulness based on the metadata (i.e., the attached online descriptions) of datasets and assigns a binary usability score. The useful datasets' descriptions are forwarded to the \texttt{Experimental Design} module.

The \texttt{Search} module thus ensures that subsequent stages of \system\ are provided with high-quality, targeted resources, establishing a robust foundation for further processes.

\subsection{Literature Processing}\label{sec:report and analyse}

\begin{figure}[h!]
    \centering
    \includegraphics[width=1\linewidth]{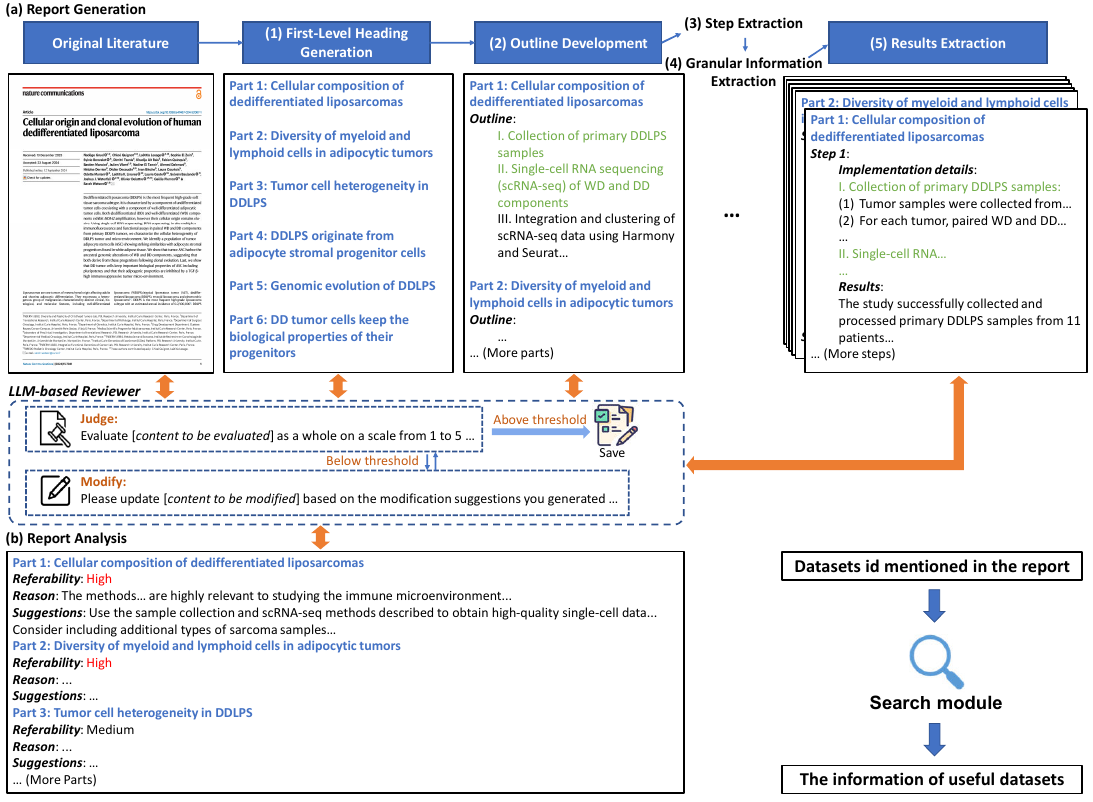}
    \caption{An example of Literature Processing. (a) illustrates the workflow for Report Generation, while (b) presents an example of Report Analysis.}
    \label{fig:literature_processing}
\end{figure}

Literature comprehension can be challenging for researchers and LLMs because the research papers are massive, lengthy, and unstructured with complex logic. To streamline literature comprehension, enhance comprehension, and provide valuable references for experimental design, we introduce the \texttt{Literature Processing} module. This module first standardizes research papers into highly structured experimental reports and analyzes them systematically. It then interacts with the \texttt{Search} module to identify the usability of the datasets mentioned in the reports. Thus, this module operates through two primary phases: report generation and report analysis. Figure~\ref{fig:literature_processing} illustrates an example of the module in operation.

\paragraph{Report Generation} Each biomedical research paper contains experiment-related contents scattered across various sections. We aim to extract and reorganize these contents into a condensed, highly structured \textit{experimental report}. Doing so brings three advantages. Firstly, the experimental report is shorter than the original paper, enhancing the efficiency of LLMs. Secondly, uniform formatting across reports provides logical structure coherence and format consistency, ensuring the LLMs grasp a big-picture idea of the most commonly acknowledged methods. Finally, the report is modularized with sections focusing on different aspects of the experiments, making it inherently suitable as a unit of analysis and a segment within retrieval-augmented generation (RAG) techniques. 

Consequently, we first introduce the hierarchical report generation process in this module, which consists of the following steps, carried out by an LLM-based report generator agent: (1) First-Level Heading Generation: The first-level headings for the experimental report are generated based on the paper's content, establishing the foundation for the report’s overall structure. 
(2) Outline Development: A comprehensive outline of the experimental report is developed, guided by the first-level headings and the paper's content, offering a structured framework for detailing experimental procedures.
(3) Step Extraction: Experimental steps are extracted from the paper and organized according to the generated outline.
(4) Granular Information Extraction: Detailed information about each experimental step is extracted from the paper. This step is critical for refining the experimental report and ensuring that all essential details are included for a thorough understanding of the experimental procedures.
(5) Results Extraction: Results related to each experimental step are extracted, facilitating the interpretation of outcomes and their relevance to the research objective.

To mitigate potential performance degradation caused by excessively long input contexts~\cite{65}, the report is divided into sections corresponding to the first-level headings, allowing parallel processing and enhancing efficiency.

\paragraph{Report Analysis} Drawing inspiration from the Chain of Thought (COT) framework~\cite{66}, which emphasizes step-by-step reasoning, we recognize that analyzing reports in relation to research objectives is a critical step in designing new experiments. Consequently, we introduce a report analysis process following report generation. Specifically, an LLM-based analyst agent evaluates each section of the experimental report for its referability, considering the research objective, conditions, and requirements. The analyst also provides suggestions for references and modifications, akin to proposing innovations on existing methods. Additionally, this module interfaces with the \texttt{Search} module to identify usable datasets within the report, using regular expressions to extract dataset IDs for retrieval from the database while bypassing the query generation step. As mentioned above, the usefulness of each dataset is determined based on its description, and useful datasets are then integrated with previously collected ones.

To ensure the quality and accuracy of the generated outputs, an LLM-based reviewer agent is introduced to interact with the experimental report generator and analyst agents at each step. Outputs are refined based on the reviewer's feedback until final approval.

To provide fine-grained retrieval that eliminates irrelevant information, upon a comprehensive review of the relevant literature, we extract and integrate the sections deemed highly referential during the analysis phase, along with their corresponding analysis content from all reports. This consolidated information serves as a reference for the \texttt{Experimental Design} module.

\subsection{Experimental Design}\label{sec:design}
\begin{figure}[h!]
  \centering
  \includegraphics[width=\textwidth]{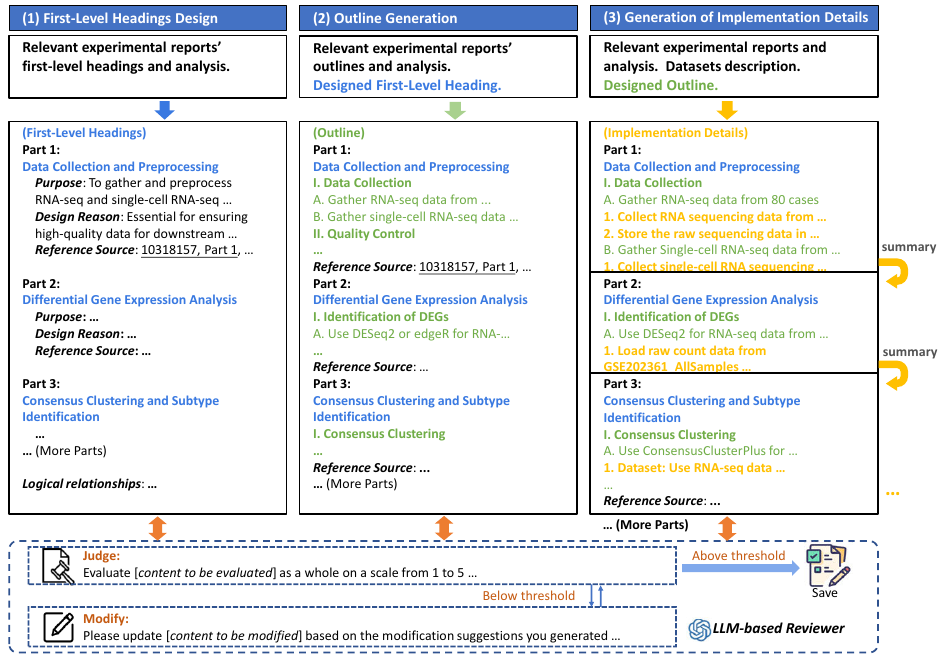}
  \caption{Demonstration of experimental design.}
  \label{fig:design case}
\end{figure}

Scientific findings in biomedical research frequently face reproducibility issues, wasting resources and time while undermining the credibility of scientific outcomes~\cite{67}. A well-designed experimental protocol is crucial for obtaining reliable results and optimizing resource use~\cite{68}. Therefore, we develop the \texttt{Experimental Design} module, which uses an LLM-based experiment designer agent to create scientific and reproducible protocols.
We propose a hierarchical learning approach to ensure the rationality of the logical structure of the generated experimental protocols. This method employs the Retrieval-Augmented Generation (RAG) technique aided by the analysis (i.e., using relevant reports' sections with high referability as reference materials), enabling the model to subsequently learn first-level headings independently, then outlines, and then experimental details from the reorganized reports. The design process is structured into three essential steps, as demonstrated in Figure~\ref{fig:design case}.

\textbf{(1) First-Level Heading Design:} The designer begins by reviewing the first-level headings and corresponding analyses of relevant experimental reports, paying particular attention to the reference and modification suggestions. Integrating this information with the research objective, conditions, and requirements, the designer crafts first-level headings for the new protocol. Additionally, for each section, the designer provides a rationale detailing the section's purpose, design reason, and reference source. This step is crucial as it lays the foundation for the experimental framework, ensuring alignment with the research objective.

\textbf{(2) Outline Generation:} The designer then constructs a brief protocol outline, referencing the outlines of pertinent reports and analyses. The designer also includes the sources of reference following the outlines. The generated outline serves as a framework for organizing the protocol.

\textbf{(3) Generation of Implementation Details:} Finally, the designer generates complete and specific implementation details for each part of the experimental protocol. Relevant sections of experimental reports and corresponding analyses are extracted based on the reference sources provided in the previous step. This information, along with the useful datasets, the protocol outline, and the summaries from earlier sections, is incorporated to produce a detailed protocol. The emphasis on detail and specificity ensures the reproducibility of the experiments, enabling researchers to follow the protocol precisely in subsequent studies.

Furthermore, an LLM-based reviewer agent, like in the Literature Processing module, is involved at each stage, providing continuous feedback to refine the design. This iterative process ensures the quality and accuracy of the final experimental design, thus contributing to the overall robustness of the research.

\subsection{Programming}\label{sec:program}
Programming in biomedical research presents a unique challenge to researchers, necessitating a combination of programming proficiency and domain-specific knowledge. To address this, we propose the \texttt{Programming} module, which is crucial in enhancing the reproducibility of experimental designs and automating systems.

To reduce the complexity of coding and debugging, the \texttt{Programming} module employs an LLM-based dry lab experiment extractor agent to derive a series of dry experiment tasks from the designed experimental protocol. Each task includes a task ID, a description of the task, and the types and descriptions of the input and output.

Subsequently, this module utilizes an LLM-based code generator agent to create R language code for each task. More programming languages will be supported in future work. To ensure code executability, a code execution within a Docker container is employed, which provides a controlled, isolated environment. Execution results-whether error reports or successful outputs-are fed back to the code generator. Based on these results, the code generator determines the next course of action: either further modification of the original code or termination of the generation process. Through this iterative cycle, the system refines the code until it achieves correctness and operational validity.

 By systematically bridging the gap between experimental design and execution, this module significantly contributes to the efficiency and efficacy of the research workflow, making it a critical module of the system's overall functionality.

\section{Evaluation Metrics}
\label{sec:metrics}
Quantitative evaluation metrics are demanded to fully reflect a research assistant's ability to advance the automation of biomedical research. However, no existing study has proposed such metrics. In this context, we propose a comprehensive method for assessing the quality of the resulting experimental protocols and programs. 

\subsection{Protocol Evaluation}
\label{sec: protocol_eval_method}
We evaluate experimental protocols from five dimensions: completeness, level of detail, correctness, logical soundness, and structural soundness. The definitions and formulas for these five metrics are as follows. 
\begin{itemize}
    \item \textbf{Completeness}: Completeness assesses how thoroughly each section of the protocol is described, considering the necessary steps that should be added to achieve the design purpose of this part. The formula for completeness is given by:
    \begin{equation}
    \text{Completeness} = \frac{\sum_{i=1}^{m} n^i_{es}}{\sum_{i=1}^{m} n^i_{ns}} = \frac{n_{ts}}{n_{ts} + n_{as}},
    \end{equation}
    where \( m \) represents the number of sections in a protocol. Here, \( n^i_{es} \) refers to the number of existing steps in the \( i \)-th section, while \( n^i_{ns} \) denotes the total necessary steps for that section. Additionally, \( n_{ts} \) indicates the total number of steps in a protocol, and \( n_{as} \) represents the number of steps that need to be added.

    \item \textbf{Level of Detail}: The level of detail measures the degree to which a protocol provides sufficient information for each step, ranging from 0 (no detail) to 1 (fully detailed).

    \item \textbf{Correctness}: Correctness assesses the proportion of protocol steps that are free from factual errors. Our analysis reveals that protocols with shorter steps tend to exhibit higher correctness scores, as the probability of factual errors increases with longer and more detailed steps. Drawing inspiration from the BLEU~\cite{25} metric in machine translation, we also introduce a brevity penalty (BP) for shorter steps. The BP is constrained to a minimum of 0.5, and the formula is as follows:
    \begin{equation}
\text{BP} = 
\begin{cases} 
1, & \text{if } l_{steps} > L, \\
max(e^{(1 - \frac{L}{l_{steps}})},~0.5), & \text{if } l_{steps} \leq L,
\end{cases}
    \end{equation}
    where $l_{step}$ represents the length of steps, quantified by the number of sentences containing more than six words within each step (this criterion is employed to exclude steps' titles from the count). The parameter $L$ denotes the average number of sentences within a step, calculated to be 4.42 from 315 protocols generated by different methods.
    
    Then, the formula for Correctness is defined as:
\begin{equation}
\label{correctness_equation}
\text{Correctness} = BP \cdot \frac{n_{cs}}{n_{ts}},
\end{equation}
where \(n_{cs}\) denotes the number of the correct steps.

    \item \textbf{Logical Soundness}: Logical soundness evaluates the proportion of steps that are placed in a reasonable order within a protocol. Reasonable steps are those that are logically ordered and appropriately positioned. The formula for Logical Soundness is given by:
\begin{equation}
\text{Logical Soundness} = \frac{n_{rs}}{n_{ts}},
\end{equation}
where \(n_{rs}\) denotes the number of reasonable steps.

\item \textbf{Structural Soundness}: Structural soundness evaluates the logical coherence and organizational integrity of a protocol's overall framework, with scores from 0 (completely unsound) to 1 (perfectly sound).
\end{itemize}

To speed up evaluation, we employ an LLM (GPT-4o) as the judge to evaluate the experimental protocols and obtain the above five metrics independently. For completeness, the LLM generates the additional steps required to achieve the design purpose of each section. For correctness and logical soundness, it evaluates the accuracy and coherence of each step. We then calculate these three metrics using their respective formulas. For the remaining two metrics—level of detail and structural soundness—the LLM directly generates the corresponding scores. Ultimately, the overall score for the experimental protocol is determined by summing the scores of all five dimensions.

\subsection{Program Evaluation}
\label{programming_eval_method}

\begin{table}[h!]
\centering
\caption{Grading criteria and error types for R code execution.}
\label{table: code_scoring_criteria}
\resizebox{\textwidth}{!}{
\begin{tabular}{clp{11cm}}
\toprule
\textbf{Level} & \textbf{Description} & \textbf{Example Error Types} \\
\midrule
1 & Minor Errors & Missing or incorrect file paths, missing necessary libraries or packages, network timeouts\\
2 & Moderate Errors & Syntax errors, incorrect function or variable names, data type mismatches\\
3 & Major Errors & Parameter mismatches, index out of bounds or invalid index, out of memory \\
4 & Severe Errors & Incorrect algorithms or logic, disorganized code structure, key components missing or incorrect \\
\bottomrule
\end{tabular}
}
\end{table}

As a critical component of research automation, the extent to which the Programming module can augment research efficiency deserves attention. As detailed in Section~\ref{sec:program}, this module generates tailored code for each dry lab experiment task. To assess its effectiveness, we propose two scoring systems. The first computes the \textbf{execution success rate}, reflecting the percentage of successfully completed tasks per protocol. The second metric assigns \textbf{error levels} to the tasks that remain incomplete, reflecting the severity of errors encountered during code execution. The detailed grading criteria are outlined in Table~\ref{table: code_scoring_criteria}. 

\section{Experiment}

In this section, we design comprehensive experiments to answer the following research questions:
\begin{enumerate}
    \item \textbf{RQ1}: How does \system~perform in automating the entire biomedical research process?
    \item \textbf{RQ2}: How does each component of \system~perform in their specific sub-tasks, including the search module (\textbf{RQ2.1}), report generation module (\textbf{RQ2.2}), and experimental design module (\textbf{RQ2.3})?
\end{enumerate}
\subsection{Experiment Setup}
\label{sec: setup}
In our experiments, we utilize the GPT-4o model as the foundational LLM for all the agents in \system. The temperature settings for various agents within the system are as follows: the query generator is configured with a temperature of 0.7 to introduce a moderate level of variability in the generated queries. Conversely, the reviewer and the LLM used for evaluation are set to a lower temperature of 0.1 to ensure more deterministic and consistent evaluations. All other agents operate at a temperature of 0.5, balancing between randomness and determinism to maintain overall coherence and reliability in the agents' outputs. Additionally, the query generator generates five queries for each user input, with a maximum retrieval quantity of 10 per query for each database. The maximum number of interaction rounds for the LLM reviewer in a single session is 6.

\paragraph{Baselines} We evaluate \system~by comparing it with three well-known agent systems. (1) \textbf{ReAct}~\cite{69}, which integrates reasoning and action within LLMs to effectively manage complex reasoning and decision-making tasks. (2) \textbf{Plan-and-Execute}~\cite{27}, which utilizes an iterative framework to accomplish tasks through sequential planning and execution. (3) \textbf{RAG}, which employs a naive Retrieval-Augmented Generation (RAG) module to search for relevant content and generates the answer in a single step. Appendix B provides a detailed description of the implementation of these baseline systems.

\subsection{Performance of End-To-End Automation}
\label{sec: end2end}
We collect eight ongoing research objectives from a biomedical laboratory to ensure that no published work has addressed these research objectives, as detailed in Appendix C (Table C1). Each objective is processed for three runs to minimize randomness. We evaluate three baseline systems and our system for designing and executing experiments for these objectives. We equip the React and Plan-and-Execute systems with four tools: (1) a search tool utilizing the NCBI API~\footnote{\url{https://www.ncbi.nlm.nih.gov/home/develop/api/}} to retrieve descriptions of relevant papers, (2) a download tool for acquiring these papers and storing them in a chunked and vectorized format, (3) a search tool using the NCBI API to obtain descriptions of relevant datasets and storing them in a chunked and vectorized format, and (4) a search tool that extracts pertinent content from the resulting vector database.

\definecolor{darkgreen}{rgb}{0.0, 0.5, 0.0}

\begin{table}[!h]
    \centering
    \caption{Quality of experimental protocols generated by various systems.}
    \label{tab: protocol}
    \resizebox{\textwidth}{!}
{ 
    \begin{tabular}{lcccccc|cc}
    \toprule
        Method & Completeness & Detail & Correctness & Logical Soundness & Structure & Overall & $l_{steps}$ & $n_{total~steps}$\\ \midrule
RAG & 0.405  & 0.687  & 0.483  & \textbf{0.973}  & \textbf{0.908}  &  3.456 &1.286 & 9.167  \\
ReAct & 0.364 & 0.577 & 0.484  & 0.963 & 0.897 & 3.285 & 1.237  &5.792 \\
Plan and Execute & 0.380 & 0.587 & 0.483  & 0.965 & 0.900 & 3.314 & 1.194 &6.375 \\ \midrule
\system~&  $\textbf{0.659}_{\uparrow\textbf{0.254}}$ & $\textbf{0.893}_{\uparrow\textbf{0.206}}$ & $\textbf{0.895}_{{\uparrow\textbf{0.411}}}$ & $0.953_{\downarrow0.020}$& $ 0.891_{\downarrow0.017}$ & $\textbf{4.292}_{{\uparrow\textbf{0.836}}}$ & 7.327 & 33.958\\\bottomrule
    \end{tabular}
    }
\end{table}

\subsubsection{Performance by Automatic Evaluation}
Table~\ref{tab: protocol} presents the average scores for protocols generated by different systems, including Completeness, Level of Detail, Correctness, Logical Soundness, and Structural Soundness. To minimize single-model evaluation bias, we employed four LLMs (GPT-4o, O3-mini-2025-01-31, Gemini-2.0-Flash, and DeepSeek-V3) for assessment, reporting their average scores across five distinct metrics. The detailed scoring of each model can be found in Appendix F (Table F1). We also calculate the overall performance, the average number of sentences per step (\(l_{\text{steps}}\)), and the average number of steps per protocol (\(n_{\text{total steps}}\)). (1) The results highlight \system's exceptional performance in Completeness, Level of Detail, and Correctness, surpassing the best baseline by 0.254 (62.7\%), 0.206 (30.0\%), and 0.411 (84.9\%), respectively. (2) Our system is comparable to the best baseline in Logical Soundness and Structural Soundness, exceeding 0.89, consistently maintaining a high performance in these aspects.
(3) Furthermore, the protocols generated by \system~have significantly more sentences per step and steps per protocol, $5.9\times$ and $4.8 \times$ greater than the average performance of different baselines, respectively, indicating the generation of more detailed and comprehensive protocols. (4) The comparative analysis of the three baselines indicates that RAG outperforms the other two iterative agent systems. This superiority can be attributed to the fact that, in the long-context environment characterized by iteration and lengthy retrieved information, the summarization and integration capabilities of the latter two systems degrade, resulting in the loss of substantial amounts of valuable information in the final generated protocol. In contrast, we design tailored workflows to effectively integrate results from multiple modules and steps, thereby preserving critical information throughout the integration process. A case comparison of protocols from the four systems is illustrated in Appendix F (Figure F1).

\begin{figure}[h!]
\centering
\begin{minipage}{0.48\textwidth}
\centering
\includegraphics[width=\textwidth]{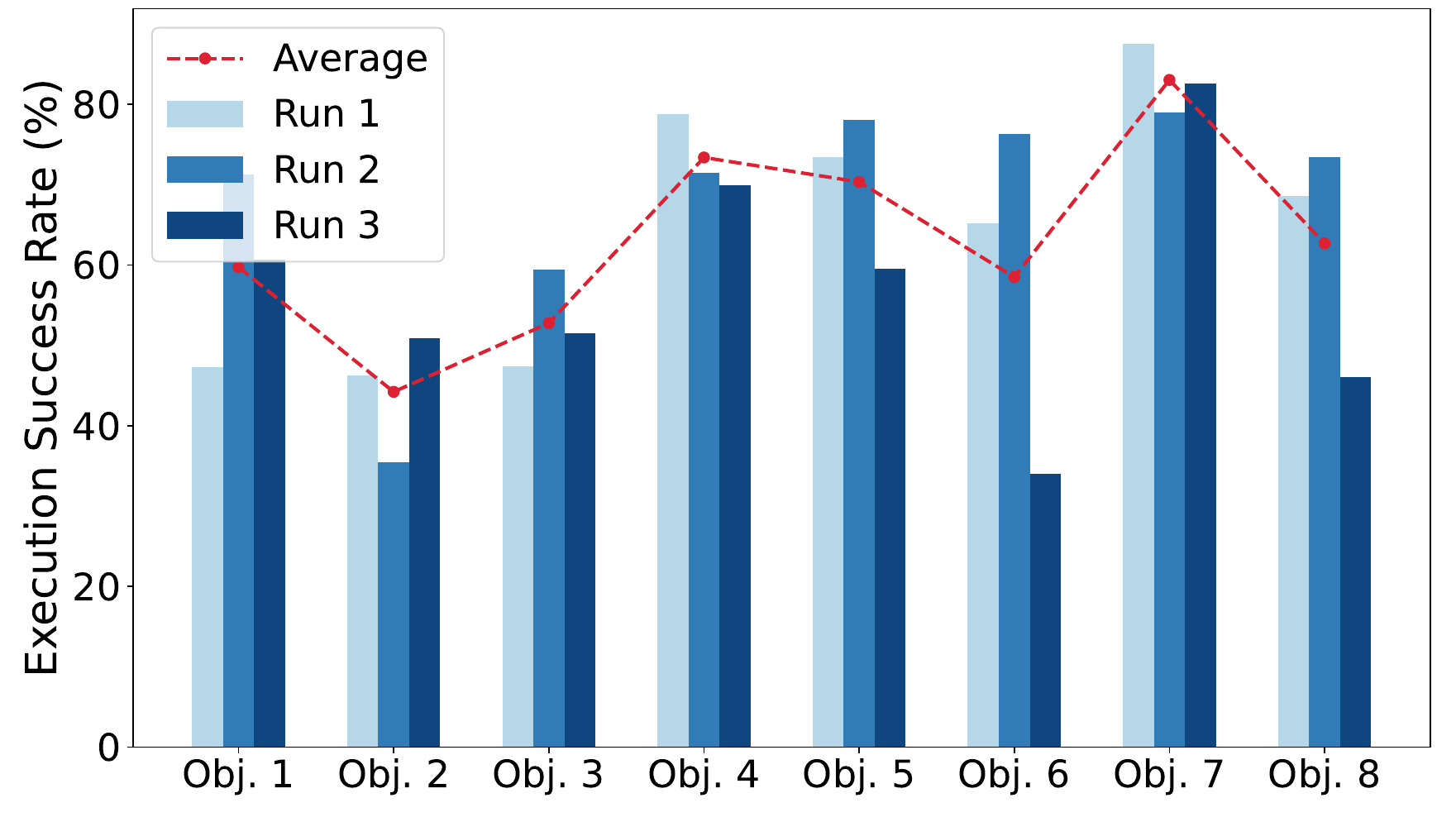}
\caption{Execution success rate.}
\label{fig:end2end success}
\end{minipage}
\begin{minipage}{0.51\textwidth}
\centering
\includegraphics[width=\textwidth]{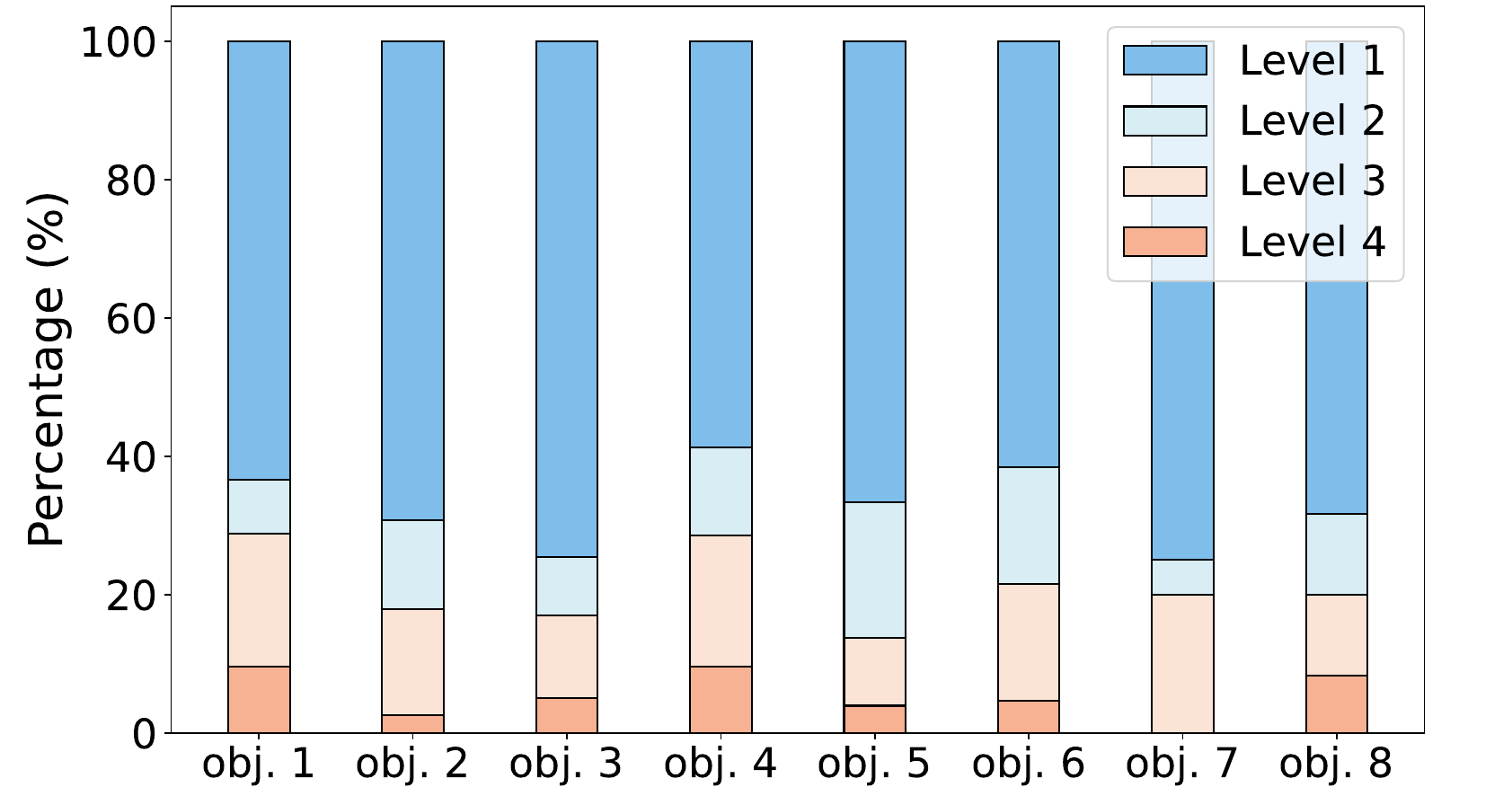}
\caption{Score frequency distribution.}
\label{fig:code2}
\end{minipage}
\end{figure}

Notably, the first three metrics are crucial for the feasibility of protocols. Consequently, in subsequent code generation and execution, our experiments reveal that the baselines almost invariably fail to produce executable code, with success rates near zero. Our approach, however, achieves an average execution success rate of 63.07\% across eight topics, with a maximum of 87.50\%, as detailed in Figure~\ref{fig:end2end success}.

Furthermore, we conduct an error analysis of tasks that failed during code execution. We assess the corresponding code using the criteria in Table~\ref{table: code_scoring_criteria} and the error messages. Figure~\ref{fig:code2} illustrates the distribution of error levels. Across eight objectives, the majority of errors are minor errors (Level 1 errors), with an average portion of 67.19\%, while severe errors requiring significant manual correction account for only 5.46\%. This indicates that even when some tasks are not executed successfully, they can be corrected with minimal human intervention.

These findings suggest that the programming module has significant potential to enhance research efficiency. While some tasks still experience errors, the majority are successfully completed without the need for human intervention. This greatly reduces the time researchers spend on coding and debugging. By further optimizing the module to reduce error rates, research automation can be improved, thereby substantially increasing overall research productivity.

\subsubsection{Quality of Automatic Evaluation}
\label{sec: human_judge_llm}

To validate the reliability of evaluations conducted by the LLM judge, we engage three domain experts to independently assess the LLM judge's evaluation outcomes. Given the time-consuming and labor-intensive nature of manual evaluation, we employed only GPT-4o, the best LLM currently, as the foundation LLM in BioResearcher to conduct multiple sampling generations on one research objective, resulting in the generation of 18 protocols used in this experiment. This focused approach also reduces evaluation complexity while enhancing assessment accuracy through deeper contextual understanding by experts.

\begin{table}[!h]
    \centering
    \caption{Performance of the LLM judge, as evaluated by human experts. Human consistency is measured using Fleiss’ kappa.}
    \label{tab: llm evaluation}
    \resizebox{\textwidth}{!}{
    \begin{tabular}{lcccccc}
    \toprule
         & Completeness & Detail & Structure & Logical Soundness & Correctness & Avg. \\ \midrule
        LLM's accuracy (\%) & 91.1 & 88.9 & 75.9 & 75.0 & 71.3 & 80.4 \\ 
        Human consistency & 0.66 & 0.92 & 0.90 & 0.88 & 0.95 & 0.86 \\
        \bottomrule
    \end{tabular}
    }
\end{table}

We engage three domain experts to systematically review the LLM judge's evaluation rationales and results for each protocol across five dimensions. This process involves verifying the factual accuracy and logical credibility of the LLM's outputs. Expert judgments are aggregated through majority voting to determine the validity of each step-level evaluation. The final analysis quantifies the LLM's assessment accuracy across five dimensions.

Results are shown in Table~\ref{tab: llm evaluation}, demonstrating that the LLM achieves an overall accuracy of 80.4\% in protocol evaluations, with exceptional performance in Completeness (91.1\%). These metrics confirm the model's capacity to deliver comprehensive and efficient automation assessments. We also calculate the inter-expert consistency using Fleiss' kappa to ensure evaluation reliability. substantial agreement is observed across all dimensions: Detail ($\kappa = 0.92$), Structure ($\kappa = 0.90$) and Correctness ($\kappa = 0.95$) reaches near-perfect consensus, while other dimensions maintain $\kappa > 0.65$. The p-values for consistency across the three dimensions are uniformly below 0.05. This agreement underscores the methodological rigor and reproducibility of the evaluation framework.

\subsubsection{Performance by Manual Evaluation}
\label{sec: human_eval_protocols}

To further validate the quality of the protocols generated by BioResearcher, we engage three domain experts to independently assess the 18 original protocols generated on one research objective, same as in Section~\ref{sec: human_judge_llm}, using consistent criteria in Section~\ref{sec: protocol_eval_method}. Concentrating on one research objective mitigates assessment complexity and enhances accuracy through deeper contextual understanding by the experts.

\begin{table}[!h]
    \centering
    \caption{Quality of experimental protocols evaluated by human experts. Consistency for the Correctness metric is measured using Fleiss' kappa, while consistency for the Detail and Structure metrics is assessed using Kendall's W.}
    \label{tab: protocol_human}
    \begin{tabular}{lcccccc}
    \toprule
         & Correctness & Detail & Structure \\
         \midrule
         Human Experts & 0.85 & 0.87 & 0.87 \\
         Human Consistency & 0.85 & 0.68 & 0.61 \\
    \bottomrule
    \end{tabular}
\end{table}

We employ experts to evaluate these protocols across three dimensions: Correctness, Detail, and Structure, following the criteria in Section~\ref{sec: protocol_eval_method}.
Specifically, the experts systematically evaluate each step of the protocols, determining whether it is correct or not to derive the \(n_{cs}\) in equation~\ref{correctness_equation} for calculating the correctness. For the remaining two metrics, the experts assign a direct score to each protocol. We exclude Completeness and Logical Soundness due to their susceptibility to subjective interpretation, as different researchers may design divergent yet valid experimental approaches. We then analyze inter-rater reliability using appropriate statistical measures. For Correctness, which quantifies the proportion of correct experimental steps in a protocol, we employ Fleiss' kappa. For Detail and Structure (ordinal scoring), we employ Kendall's W. As shown in Table~\ref{tab: protocol_human}, the results in three dimensions are all over 85\%, and all consistency metrics exceed 0.6, with Correctness over 0.8. The p-values for consistency across the three dimensions are uniformly below 0.05. These results confirm strong consensus among evaluators across all dimensions.

\subsection{Performance of Search Module}
\label{sec: search_eval}
\subsubsection{The Effect of Generated Queries} 
We conduct a comparative experiment to assess the effectiveness of LLM-generated queries. An LLM and three human participants independently generate five queries for each user input, which specifies a research objective, conditions, and requirements. The evaluation metric is the number of relevant papers or datasets retained after retrieval and filtering. The LLM generates queries three times, with the results averaged, and a similar average is calculated across the three human participants. This experiment, covering ten research objectives listed in Appendix C (Table C2), presents its results in Figure~\ref{fig:human_vs_llm}.

\begin{figure}[h]
  \centering
    \includegraphics[width=\linewidth]{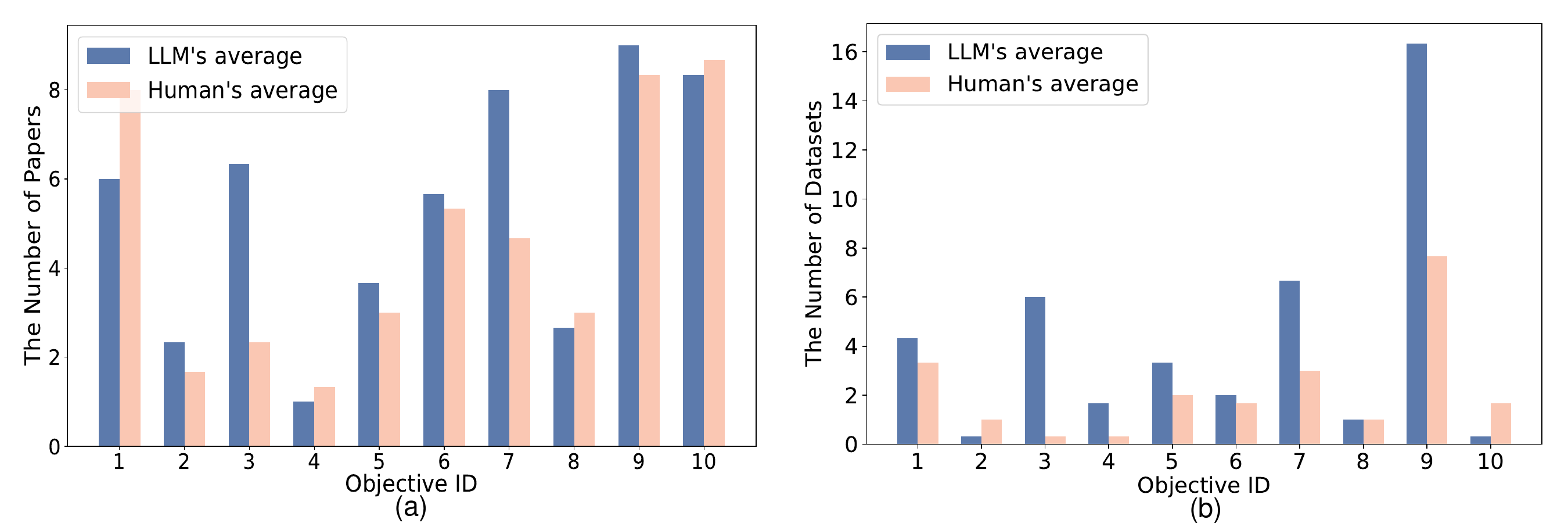}
  \caption{Comparison of the effects of LLM-generated queries and human-generated queries. (a) Comparison of the number of useful papers retrieved by LLM- and human-generated queries. (b) Comparison of the number of useful datasets retrieved by LLM- and human-generated queries.}
  \label{fig:human_vs_llm}
\end{figure}

LLM-generated queries generally outperform human-generated ones across most objectives. (1) In Figure~\ref{fig:human_vs_llm}(a), which compares the number of useful papers retrieved, the LLM-generated queries show significantly stronger performance in objectives 3 and 7, but the increases are modest on objectives 2, 5, 6, and 9. Human-generated queries perform slightly better in three objectives, suggesting that human intuition can offer an edge in certain cases. (2) Figure~\ref{fig:human_vs_llm}(b) presents a similar pattern in dataset retrieval. The LLM outperforms human participants in retrieving useful datasets for eight objectives, particularly Objective 3, where the LLM retrieved an average of six datasets compared to the human average of one-third. This indicates that only one human participant successfully generated queries that retrieved one useful dataset. Conversely, for objectives such as Objective 2 and Objective 10, human-generated queries show a slight advantage. 

However, a key advantage of LLMs lies in their efficiency and scalability. Unlike human participants, who may take longer to generate queries and may experience fatigue, LLMs can quickly generate multiple queries with minimal effort. Furthermore, LLMs can mitigate performance gaps in challenging objectives by repeating query generation to increase the chances of retrieving relevant papers and datasets. This iterative capability allows LLMs to adapt and improve results, making them highly effective for large-scale or repetitive search tasks.

\subsubsection{The Effect of LLM-based Filter Agent} To assess the precision of the ratings assigned by the LLM-based filter agent, we engage human reviewers to undertake a parallel assessment using the scoring criteria outlined in Table ~\ref{table: scale}. Both the filter agent and human reviewers evaluate the same set of papers and datasets, with their ratings based solely on each paper's title and abstract or the dataset's description. We use Kendall’s W to assess agreement for the ordinal ratings of 150 papers, as it is suitable for measuring concordance in ordinal data. For the binary ratings of 50 datasets, we apply Fleiss' kappa, which evaluates inter-rater agreement for categorical data among multiple raters. These papers and datasets are sampled from the search results of 10 topics listed in Table~\ref{tab:search topics}. Higher values in both metrics indicate stronger agreement, enhancing the reliability of the human ratings. A majority voting mechanism establishes the ground truth for human ratings, which serves as the benchmark for calculating the model's accuracy. For papers, those with scores of 4 or higher are classified as useful, while others are deemed not useful, and the accuracy is calculated based on this binary classification. The results of this study are represented in Table~\ref{tab:reviewer_consistency}.

\begin{table}[h]
\centering
    \caption{Evaluation results of the LLM-based filter agent. Human consistency in paper reviews is measured by Kendall's W, and in dataset reviews, Fleiss' kappa is employed. All p-values are less than 0.05, indicating statistical significance. LLM accuracy is calculated using the ground truth from the majority of human ratings.}
    \label{tab:reviewer_consistency}
\begin{tabularx}{\textwidth}{l|XXXXXXXXXXX}
\toprule
Objective ID & 1 & 2 & 3 & 4 & 5 & 6 & 7 & 8 & 9 & 10 & Total \\\midrule
 & \multicolumn{11}{c}{Papers Review} \\\midrule
Human's consistency & 0.67 & 0.80 & 0.84& 0.91 & 0.86 & 0.80 & 0.75 & 0.75 & 0.80 & 0.68 & 0.82 \\
LLM's   Accuracy & 80\% & 73\% & 80\% & 93\% & 93\% & 93\% & 80\% & 80\% & 87\% & 80\% & 84\% \\\midrule
 & \multicolumn{11}{c}{Datasets Review} \\\midrule
Human's   consistency &  0.66&  0.7&  0.66&  1.00&  0.70&  0.70&  1.00&  1.00&  1.00&  0.73& 0.84 \\
LLM's   Accuracy & 80\% & 80\% & 80\% & 80\% & 60\% & 80\% & 100\% & 80\% & 100\% & 100\% & 84\%\\
\bottomrule
\end{tabularx}
\end{table}

The results indicate strong consistency among human reviewers and notable accuracy of the LLM-based filter. For paper reviews, human consistency, measured by Kendall’s W, shows an overall concordance of 0.82, reflecting substantial agreement. Based on the established human ground truth, the filter achieves an average accuracy of 84\% in classifying papers as useful or not. The filter's accuracy peaks at 93\% for several objectives, underscoring its effectiveness in aligning with human assessments. In dataset reviews, human consistency, assessed using Fleiss' kappa, exhibits strong agreement with a value of 0.84. The filter maintains an average accuracy of 84\%, with perfect accuracy in specific objectives, highlighting its reliability in dataset classification. The statistical significance of these results, confirmed by p-values less than 0.05, reinforces the robustness of the filter in achieving high concordance with human evaluations across both domains.

\subsection{Performance of Report Generation}
\label{sec: report_eval}

\begin{figure}
    \centering
    \includegraphics[width=0.5\linewidth]{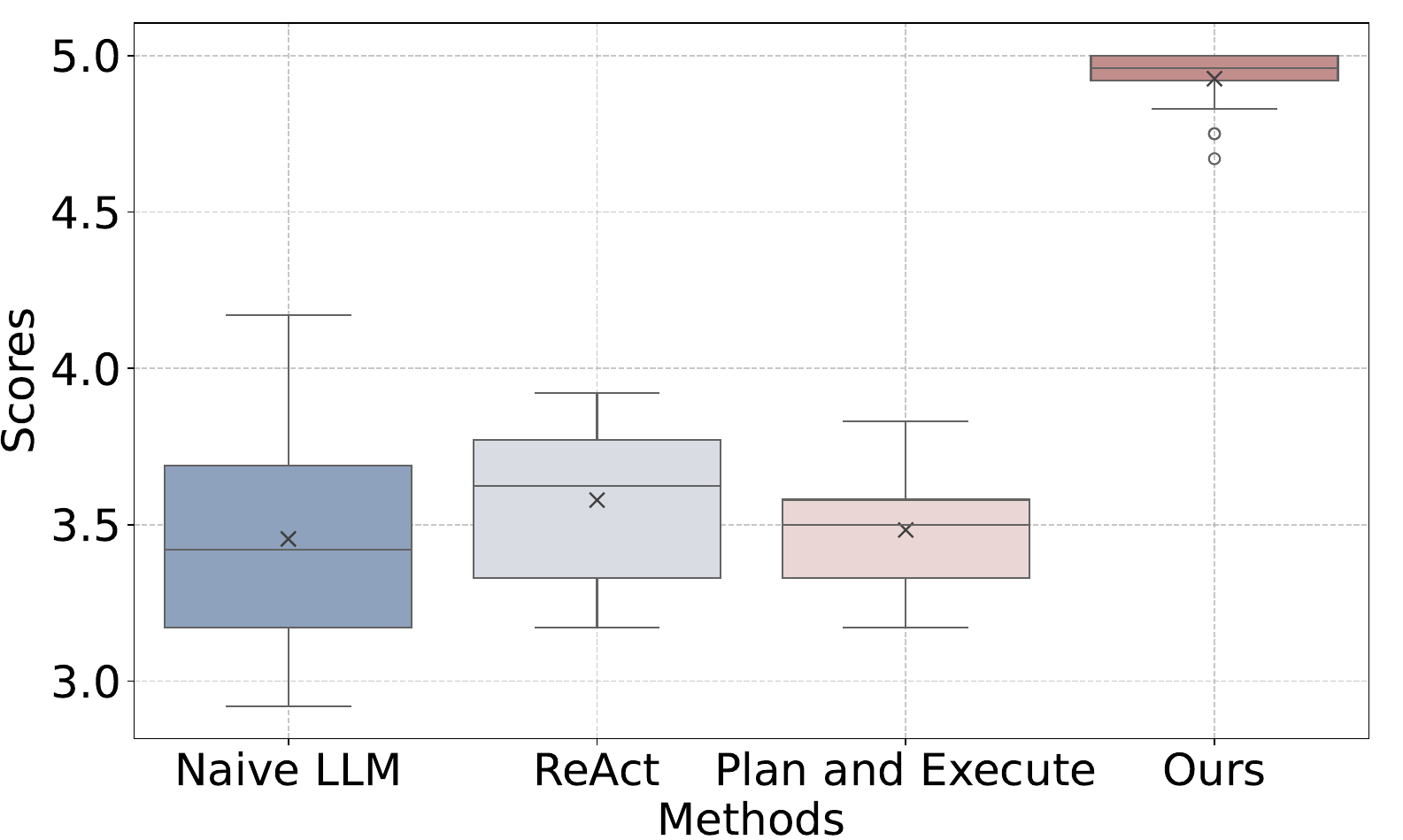}
    \caption{Boxplot of scores for experimental reports generated by different methods.}
    \label{fig:report}
\end{figure}

To test the impact of our hierarchical report generation method, we compare its reports against those produced by ReAct, Plan-and-Execute, and a naive single-step LLM approach. Specifically, we standardized 20 papers into an experimental report format using each method. None of the methods are equipped with any additional tools. We prompt an LLM to evaluate the reports across four dimensions: logical soundness, level of detail, consistency with the original paper, and readability. The model assigns a score ranging from 1 to 5 for each dimension, with the scoring criteria detailed in Appendix D (Table D2). The final score is calculated as the average across these four dimensions. 
Figure~\ref{fig:report} presents a box plot of the scores for reports generated by the four different methods.

As illustrated in Figure~\ref{fig:report}, our method consistently outperforms the others, demonstrating significantly higher median values and narrower interquartile ranges, indicative of both superior performance and stability. Conversely, the three baseline models fail to generate high-quality reports. Among them, the Naive LLM shows the greatest score variability, ranging from 2.92 to 4.17, underscoring its instability. In contrast, the results for ReAct and Plan-and-Execute are concentrated around 3.5 points and do not exceed 4 points, reflecting their limited potential. We attribute the poor performance of the baselines to their failure to generate high-quality, long-length outputs when handling the extensive context of an entire paper. Our method, employing a hierarchical approach, enables the model to iteratively generate shorter outputs, which are then integrated into coherent, high-quality reports. These detailed, accurate, and well-structured reports provide excellent references for subsequent experimental design.

\subsection{Performance of Experimental Design}
\label{sec: design_eval}
To validate the efficacy of the experimental design module, we construct a comparison experiment against the three baselines introduced in Section~\ref{sec: setup}.  We employ the search module and Literature Processing module of \system~for 15 research objectives listed in Appendix C (Table C3), thereby obtaining reports and analyses that serve as a knowledge base. The three baselines can invoke search tools to retrieve relevant content from this knowledge base. All experiments are repeated three times, and the evaluation results are averaged and presented in Table~\ref{tab: design}.

\begin{table}[!h]
    \centering
    \caption{Scores of experimental protocols generated by various methods.}
    \label{tab: design}
\resizebox{\textwidth}{!}
{ 
    \begin{tabular}{lcccccc|cc}
    \toprule
        Method  &Completeness & Detail & Correctness & Logical Soundness & Structure & Overall  &$l_{steps}$ & $n_{total~steps}$ \\ \midrule
RAG & 0.451 & 0.843 & 0.506 & 0.970 & \textbf{0.941} & 3.711 & 2.005 & 10.178\\
ReAct &  0.445 & 0.778 & 0.516 & 0.972 & 0.926 & 3.636 &1.949 & 9.244\\
Plan and Execute & 0.388 & 0.727 & 0.544 & 0.965 & 0.920 & 3.544 &1.987 & 8.133\\\midrule
Our Method & \textbf{0.582} & \textbf{0.901} & \textbf{0.946} & \textbf{0.979} & 0.913 & \textbf{4.321} & 6.477 &15.289\\ 

\bottomrule
    \end{tabular}
    }
\end{table}

 The results in Table~\ref{tab: design} show that our methodology surpasses the three baseline methods in all metrics except Structural Soundness. Specifically, our approach exhibits superior performance in Completeness, Level of Detail, Correctness, and Logical Soundness, exceeding the best baseline scores by 0.131 (29.0\%), 0.058 (6.9\%), 0.402 (73.9\%), and 0.007 (0.7\%), respectively. This indicates a more comprehensive, detailed, accurate, and logically sound generation of experimental protocols. Although our method does not lead in the Structural Soundness metric, it still achieves a commendable score of 0.913, and its overall performance is the highest among all methods. These findings confirm that our hierarchical learning approach, which incrementally designs experimental protocols, effectively addresses the challenges posed by lengthy inputs and outputs, resulting in higher-quality experimental protocols.

Referring to Table~\ref{tab: protocol}, we observe an improvement in the average performance of all three baseline methods, particularly in Completeness (0.373 to 0.428), Level of Detail (0.666 to 0.783), and Correctness (0.481 to 0.522). Their average \( l_{\text{steps}} \) and \( n_{\text{total steps}} \) also increase, with the averages rising from 1.239 to 1.980 and 7.111 to 9.185, respectively. The complexity of the research objectives discussed in this section is consistent with those in Table~\ref{tab: protocol}, rendering this comparison meaningful. It also underscores the significant role of our report generation in reducing irrelevant information and providing a valuable reference for experimental design.
\begin{table}[!h]
    \centering
    \caption{Scores of experimental protocols generated by \system~without experimental report analyst or reviewer.}
    \label{tab: ablation}
\resizebox{\textwidth}{!}
{ 
    \begin{tabular}{lcccccc|cc}
    \toprule
        Method  &Completeness & Detail & Correctness & Logical Soundness & Structure & Overall  &$l_{steps}$ & $n_{total~steps}$ \\ \midrule
Our Method & 0.582 & 0.901 & 0.946 & 0.979 & 0.913 & 4.321 & 6.477 &15.289\\ 
w/o Reviewers & 0.550 & 0.884 & 0.901 & 0.970 & 0.918 & 4.223 & 4.583 & 15.333\\
w/o Analyst & 0.559 & 0.901 & 0.947 & 0.982 & 0.919 & 4.308 & 6.586 & 14.978\\
\bottomrule
    \end{tabular}
    }
\end{table}

Additionally, we assess the role of the LLM Reviewers in quality control for the automation of biomedical research. Removing the reviewers from both the literature processing and experimental design modules leads to a decline in performance across most metrics, as shown in Table~\ref{tab: ablation}. Specifically, the inclusion of the LLM Reviewers results in an overall score increase of 0.098, with the most notable improvement of 0.045 in Correctness. This indicates that the LLM reviewers effectively identify and correct errors in literature processing and experimental design, thereby enhancing the accuracy of the output. Moreover, our system with reviewers generates more detailed and comprehensive protocols, as evidenced by higher scores in Completeness, Level of Detail, and the average number of sentences per step.

To evaluate the impact of the report analyst agent within the Literature Processing module, we remove this agent from \system. Instead, we use the same retrieval model employed in the baselines to extract relevant content from reports as references for protocol generation. As shown in Table~\ref{tab: ablation}, the Completeness score significantly declines from 0.582 to 0.559. This drop is due to the analyst providing specific references and modification suggestions that enhance protocol completeness. However, in the other four dimensions, removing the report analyst does not negatively affect outcomes and even results in slight improvements. These minor improvements, averaging an increase of only 0.0025, can be considered normal fluctuations. Furthermore, from the perspectives of system interpretability and user-friendliness, the report analyst helps users understand how the system designs new experimental protocols based on relevant materials.

\section{Conclusion}\label{sec:conclusion}
In this study, we introduced \system, an intelligent research assistant that automates the biomedical research process. Utilizing a modular LLM-based multi-agent architecture, \system~addresses the multidisciplinary demands, logical complexities, and performance evaluation challenges of biomedical research. It automates tasks such as literature review, experimental protocol design, and code implementation, significantly improving research efficiency and reducing manual workload. We developed novel evaluation metrics focusing on protocol quality and experimental automation, providing a robust framework for assessing performance. Our results show that \system~designs executable experimental protocols with a high success rate, outperforming existing typical agent systems.

The practical significance of \system~lies in its ability to automate the research pipeline, allowing researchers to focus on strategic decision-making and innovation. This advancement accelerates biomedical discoveries and future developments in automated research systems. By potentially extending its capabilities to wet lab experiments, \system~promises broader applications. This study lays the groundwork for enhancing automated research technologies, contributing to global health and scientific progress.

\Acknowledgements{Yeyun Gong and Chen Lin are the corresponding authors. Chen Lin is supported by the Natural Science Foundation of China (No. 62432011) and Zhongguancun Academy (No. 20240310).}

\Supplements{Appendix A-F.}

\Citation{
If you find our work useful in your research, please cite us using the following BibTeX:\\

@article\{BioResearcher,\\
author = "Yi Luo and Linghang Shi and Yihao Li and Aobo Zhuang and Yeyun Gong and Ling Liu and Chen Lin",\\
title = "From Intention To Implementation: Automating Biomedical Research via LLMs",\\
journal = "SCIENCE CHINA Information Sciences",\\
year = "2025"\\
\}
}


\newpage


\begin{appendix}

\section{Limitations and Future Works}
\paragraph{Limitations}

In this paper, \system~effectively automates the entire biomedical research process, from literature and dataset searches to experimental design and execution, given a research objective and specified conditions. However, several limitations warrant discussion. First, the system does not achieve complete success in executing experiments without manual intervention. This limitation is partly due to the need for further enhancement of the code generator agent's performance. Second, certain anomalies, such as the unavailability of resources specified in the experimental protocols, also hinder fully automated execution. It highlights the necessity of anticipating a broader range of exceptional scenarios and developing corresponding solutions in future work. Third, during our practice, we observed that excessively long inputs often result in LLMs producing shorter, more generalized outputs. To enable LLMs to generate detailed and comprehensive experimental protocols, we employed the multi-step, section-wise processing approach at various stages, including experimental report generation and experimental design. However, this iterative method increases the overall cost. Fourth, the framework currently does not formally model the research process as a constrained optimization problem. This limits its ability to optimally balance competing objectives (e.g., accuracy vs. computational cost) or dynamically allocate resources (e.g., lab equipment, cloud computing budgets), particularly in complex, multi-agent research scenarios.

\paragraph{Future Works}

To further enhance and expand the capabilities of \system, we propose four key future research directions. First, \system~currently supports only dry lab experiments. Future research could explore integrating automated wet lab technologies, such as Cloud Labs~\cite{70}, to extend the system's applicability to wet lab studies. Additionally, its utility could be expanded to other scientific disciplines through three modular adaptations: (1) Extensible code modules supporting additional languages (Python/Julia) and domain-specific tools (LabVIEW for physics experiments); (2) Customizable knowledge bases using field-specific repositories (arXiv for CS/physics, ChemRxiv for chemistry); (3) Specialized innovation modules employing generative AI for hypothesis generation in creativity-driven domains. Second, hallucination remains a significant challenge in LLM-based applications, where the model may generate factually incorrect content. This issue is particularly critical in scientific research due to the model's limited domain-specific knowledge. Addressing LLM hallucination continues to be an essential area for future work. Third, incorporating human oversight mechanisms is crucial for ensuring reliability and ethical compliance in complex scientific workflows. We propose three avenues for integrating human intervention: (1) Confidence-Driven Human Intervention, where a confidence threshold mechanism prompts human validation when the reviewer agent's confidence falls below a predefined threshold; (2) Iterative Refinement via Natural Language Feedback, allowing users to provide corrections (e.g., missing controls or outdated references), which the AI agents will process and update accordingly while maintaining audit trails; and (3) Ethical Decision-Making Protocol, introducing an automated ethics checkpoint that flags high-risk proposals, mandates human approval for sensitive experiments, and integrates institutional ethics guidelines into the system's constraints. Future work will quantitatively evaluate the trade-offs between automation and human intervention to optimize system performance and usability. Fourth, incorporating constrained optimization techniques would enable resource-aware scheduling algorithms that balance computational budgets, equipment availability, and temporal constraints through multi-objective optimization -- particularly valuable for large-scale collaborative studies.

\section{Details of Baselines Implementations}
\label{sec:baselines_implement}
In Section 5.2, due to the NCBI databases' limitations in semantic similarity searches, directly using the user's input objective poses challenges in retrieving relevant papers and datasets. Thus, we provide the RAG system with unfiltered papers and datasets from our Search module, enabling it to extract relevant datasets and content to generate a comprehensive protocol and corresponding code in a single step.

In Section 5.4, we directly supply the literature to be processed to the three systems, thereby not equipping them with any tools. 

In Section 5.5, we equip the three systems with the aforementioned fourth tool. Here, we chunk each report and analysis by parts, rather than using the semantic similarity approach employed for chunking the papers in Section 5.2, and then store them in a vectorized format for the search tool.

We implement all baselines using the AutoGen framework~\cite{71} and adopt OpenAI's text-embedding-ada-002 as the embedding model in RAG applications.

\section{Details of Research Objectives}
In this section, we outline the specific research objectives utilized in our experiments. Table~\ref{tab:topics} presents eight objectives employed for end-to-end automation experiments. These objectives are previously unmet and are gathered from a biomedical research laboratory. Table~\ref{tab:search topics} displays ten objectives sourced from publicly available papers for evaluating the search module. Table~\ref{tab: design topics} presents fifteen objectives for assessing experimental design; these are also collected from published papers, but their publication dates are later than the cutoff for GPT-4o training data.
\begin{table}[h!]
    \centering
    \caption{8 research objectives used for end-to-end automation experiment.}
    \label{tab:topics}
\begin{tabularx}{\textwidth}{lp{15cm}}
    \toprule
       ID & Research Objective\\\midrule
        1 & Classification of the Immune Microenvironment and Identification of Key Genes in Liposarcoma Based on Transcriptomics.
\\
        2 & RNA-seq-based analysis of differences in metabolic characteristics between Well-differentiated and dedifferentiated Liposarcoma.
\\
        3 & Comprehensive analysis of gene expression and immune microenvironment differences between retroperitoneal and limb liposarcoma.
\\
        4 & Prognostic and Gene Expression Differences between Well-Differentiated and dedifferentiated Liposarcoma.
\\
        5 & Differences in molecular expression and microenvironment between primary and recurrent liposarcoma.
\\
        6 & Study on the recurrence mechanism of liposarcoma based on transcriptome and establishment of a prediction model.
\\
        7 & Key Genes and Pathways for Evaluating the Efficacy of MDM2 Inhibitor Therapy in Liposarcoma.
\\
        8 & Critical Genes and Pathways Associated with Radiotherapy Sensitivity in Retroperitoneal Well-Differentiated and Dedifferentiated Liposarcoma.
\\\bottomrule
    \end{tabularx}
\end{table}

\begin{table}[h!]
    \centering
    \caption{10 research objectives used in the evaluation of the Search module.}
    \label{tab:search topics}
\begin{tabularx}{\textwidth}{lp{15cm}}
    \toprule
       ID & Research Objective\\\midrule
        1 &  Identify populations that are sensitive to immunotherapy for liposarcoma and explore the mechanisms of therapeutic sensitivity to guide the treatment of those whose immune quality is not sensitive.
\\
        2 &  To investigate the expression characteristics of the TBC1D4 activating protein molecule and identify key module genes for preventing the progression of thyroid cancer. Through bioinformatics analysis, elucidate the role of TBC1D4 in thyroid cancer and its regulated gene network to provide new molecular targets and research directions for early prevention and treatment of thyroid cancer.
\\
        3 & To explore malignant cell characteristics related to microvascular invasion (MVI) in hepatocellular carcinoma (HCC) through multi-omics transcriptomic analysis and develop a machine learning prognostic model based on MVI-related genes.
\\
        4 & To develop a machine learning classifier based on host immune response mRNA to accurately distinguish between acute viral respiratory infections (viral ARI) and non-viral respiratory infections (non-viral ARI). Identify and validate gene expression characteristics that can aid in clinical diagnosis using host gene expression data from nasal swab samples through a multi-omics analysis framework.
\\
        5 & To identify preventive biomarkers related to DNA replication in ovarian cancer through gene expression analysis and bioinformatics analysis, particularly focusing on the expression of MCM2 protein and its role in ovarian cancer progression.
\\
        6 & To analyze coagulation-related genes in hepatocellular carcinoma (HCC) and explore their relationship with the tumor immune microenvironment (TME) and clinical prognosis.
\\
        7 & To systematically analyze long non-coding RNAs (lncRNAs) related to immune checkpoints (ICP) and explore their functions in cancer, as well as their potential as biomarkers for predicting immune therapy responses and prognosis.
\\
        8 &  To reveal the transcriptional patterns of HER2 through a pan-cancer analysis of HER2 indices, facilitating a more precise selection of patients suitable for HER2-targeted therapy.
\\
        9 & To construct an immune-related gene prognostic index (IRGPI) to predict the prognosis of head and neck squamous cell carcinoma (HNSCC) patients and clarify the molecular and immune characteristics of different HNSCC subgroups defined by IRGPI, as well as the efficacy of immune checkpoint inhibitor (ICI) therapy.
\\
        10 & To reveal the heterogeneity of T cell exhaustion (TEX) in the tumor microenvironment (TME) of different cancer types, depict the hierarchical functional dysfunction process of T cell exhaustion through pan-cancer analysis and investigate its association with prognosis and immune therapy efficacy.
\\\bottomrule
    \end{tabularx}
\end{table}

\begin{table}[h!]
    \centering
    \caption{15 research objectives used in the evaluation Experimental Design.}
    \label{tab: design topics}
\begin{tabularx}{\textwidth}{lp{15cm}}
    \toprule
       ID & Research Objective\\\midrule
        1 & Analysis of differential expression characteristics of TBC1D4 gene in different stages of thyroid cancer progression.
\\
        2 & Construction of TBC1D4-related gene co-expression modules and functional annotation.
\\
        3 & Identification of differentially expressed genes in thyroid-related lesions.
\\
        4 & Screening of core genes in key modules of thyroid cancer based on LASSO algorithm.
\\
        5 & Identification of malignant cell subpopulations in hepatocellular carcinoma by single-cell RNA sequencing.
\\
        6 & Pseudo-time analysis to explore the differentiation trajectory of malignant cells in hepatocellular carcinoma.
\\
        7 & Exploring the communication pathways of cells related to microvascular invasion in hepatocellular carcinoma based on intercellular communication analysis.
\\
        8 &  Using GO and KEGG pathway analysis to reveal the functional enrichment of MCM2-related differentially expressed genes in ovarian cancer.
\\
        9 & Analysis of the relationship between the expression level of MCM2 gene in ovarian cancer and clinical prognosis.
\\
        10 & Analysis of the functional status of CD4+ T cell subsets in rheumatoid arthritis.
        \\
        11 & Analysis of key immune cell type composition characteristics in pancreatic ductal adenocarcinoma immune cell death-related gene subtypes.
        \\
        12 & Analysis of immune checkpoint expression patterns in high and low subtypes of immune cell death-related genes in pancreatic ductal adenocarcinoma.
        \\
        13 & Analysis of tumor microenvironment characteristics of pancreatic ductal adenocarcinoma based on high and low expression subtypes of immune cell death-related genes.
        \\
        14 & Screening of potential anti-tumor drugs based on the expression pattern of ribosome production genes.
        \\
        15 & Identification of ribosome production genes with high-frequency copy number variation in cancer.
\\\bottomrule
    \end{tabularx}
\end{table}

\section{Scoring Criteria}
In this section, we outline the specific scoring criteria used in our experiments. Table~\ref{table: scale} presents the criteria for evaluating a paper's helpfulness in achieving a specific research objective. Additionally, Table~\ref{tab:report_criteria} displays the criteria for assessing the quality of an experimental report across four dimensions.

\begin{table}[h!]
    \centering
        \caption{Scoring criteria for the helpfulness of papers.}
        \label{table: scale}
        \begin{tabular}{cp{14cm}}
            \toprule
            \textbf{Rating} & \textbf{Description} \\ \midrule
            1 & \textbf{Not Helpful at All}: \\& -The title and abstract of the paper are completely unrelated to the research objective, conditions, and requirements. -Provides no valuable information or insights. \\ \midrule
            2 & \textbf{Slightly Helpful}: \\ & -The title and abstract of the paper have some minor relevance to the research objective, conditions, and requirements. -Provides very limited information and insights, making it difficult to contribute significantly to the research objective. \\ \midrule
            3 & \textbf{Moderately Helpful}: \\ & -The title and abstract of the paper are somewhat relevant to the research objective, conditions, and requirements. -Provides some useful information and insights, but additional information or research may be needed to fully utilize it. \\ \midrule
            4 & \textbf{Very Helpful}: \\ & -The title and abstract of the paper are highly relevant to the research objective, conditions, and requirements. -Provides a substantial amount of useful information and insights, making a significant contribution to the research objective. \\ \midrule
            5 & \textbf{Extremely Helpful}: \\ & -The title and abstract of the paper perfectly align with the research objective, conditions, and requirements. -Provides critical information and insights that directly and significantly advance the research objective. \\ \bottomrule
        \end{tabular}
\end{table}

\begin{table}[h!]
\centering
\caption{Scoring criteria for experimental reports.}
\label{tab:report_criteria}
\begin{tabular}{p{0.3\textwidth} p{0.7\textwidth}}
\toprule
\textbf{Dimension} & \textbf{Criterion} \\
\midrule

\textbf{Logical Soundness} & 
\begin{tabular}[t]{@{}p{0.6\textwidth}@{}}
5: Completely logical arguments and conclusions, clear reasoning, no gaps. \\
4: Mostly logical, minor unclear points or leaps. \\
3: Generally logical, some noticeable gaps or inconsistencies. \\
2: Many logical flaws, disorganized logic. \\
1: Lacks coherence, severe flaws undermining arguments. \\
\end{tabular} \\

\midrule

\textbf{Detail Level} & 
\begin{tabular}[t]{@{}p{0.6\textwidth}@{}}
5: All necessary details, comprehensive descriptions. \\
4: Most necessary details, some areas brief. \\
3: Some necessary details, lacks important information. \\
2: Lacks many details, unclear descriptions. \\
1: Almost no details, difficult to understand. \\
\end{tabular} \\

\midrule

\textbf{Consistency with Original Paper} & 
\begin{tabular}[t]{@{}p{0.6\textwidth}@{}}
5: Entirely faithful, no misunderstandings. \\
4: Mostly faithful, minor misunderstandings. \\
3: Generally faithful, noticeable misunderstandings. \\
2: Many inconsistencies, significant deviations. \\
1: Severely inconsistent, obvious misunderstandings. \\
\end{tabular} \\

\midrule

\textbf{Readability of Report Structure} & 
\begin{tabular}[t]{@{}p{0.6\textwidth}@{}}
5: Clear structure, logical organization, easy to read. \\
4: Mostly well-structured, minor section issues. \\
3: Generally clear, some poorly arranged sections. \\
2: Quite disorganized, poor reading experience. \\
1: Very disorganized, extremely difficult to read. \\
\end{tabular} \\

\bottomrule
\end{tabular}
\end{table}

\section{Cost Analysis}

Below we illustrate the average number of tokens and the associated costs involved in generating an experimental protocol and its corresponding R code based on specified research targets, conditions, and requirements, using various large language models (LLMs) in Table~\ref{tab: cost_analysis}. The process also encompasses the execution and verification of the code. In addition, after investigation, for a biomedical research cycle from the research target to the final dry experimental results, we compared the approximate manual time obtained from our survey with the time consumed by BioResearcher in Table~\ref{tab: time_analysis}.

\begin{table}[h!]
    \centering
    \caption{Cost of \system}
    \label{tab: cost_analysis}
    \begin{tabular}{cccccccc}
    \toprule
 Phase & Input Toktens & Output Toktens & GPT-4o-mini & GPT-4o-2024-08-06 & GPT-4o cost \\ \midrule
 Search & 0.09M & 0.02M & 0.03\$ & 0.45\$ & 0.91\$ \\
 Literature Processing & 49.62M & 1.74M & 8.49\$ & 141.48\$ & 282.96\$ \\
 Design & 0.77M & 0.06M & 0.15\$ & 2.48\$ & 4.96\$ \\
 Program & 61.08M & 2.57M & 10.70\$ & 178.39\$ & 344.07\$ \\ \midrule
 Total & 111.55M & 4.39M  & 19.37\$ & 322.81\$ & 632.90\$ \\
\bottomrule
    \end{tabular}
\end{table}

\begin{table}[h!]
    \centering
    \caption{Time comparison between human and \system}
    \label{tab: time_analysis}
    \begin{tabular}{cccccccc}
    \toprule
    \textbf{Phase} & \textbf{Human} & \textbf{BioResearcher} \\ \midrule
    Search & 1-3 Weeks & 0.37 hours \\
    Literature Processing & 1-2 Weeks & 5.03 hours \\
    Design & 1 Weeks & 0.16 hours \\
    Program & 4-8 Week & 2.6 hours \\ \midrule
    \textbf{Total} & \textbf{7-14 Weeks} & \textbf{8.16 hours} \\
    \bottomrule
    \end{tabular}
\end{table}

\section{Comparison of protocols generated by 4 systems.}
\begin{figure}[h!]
  \centering
  \includegraphics[width=\textwidth]{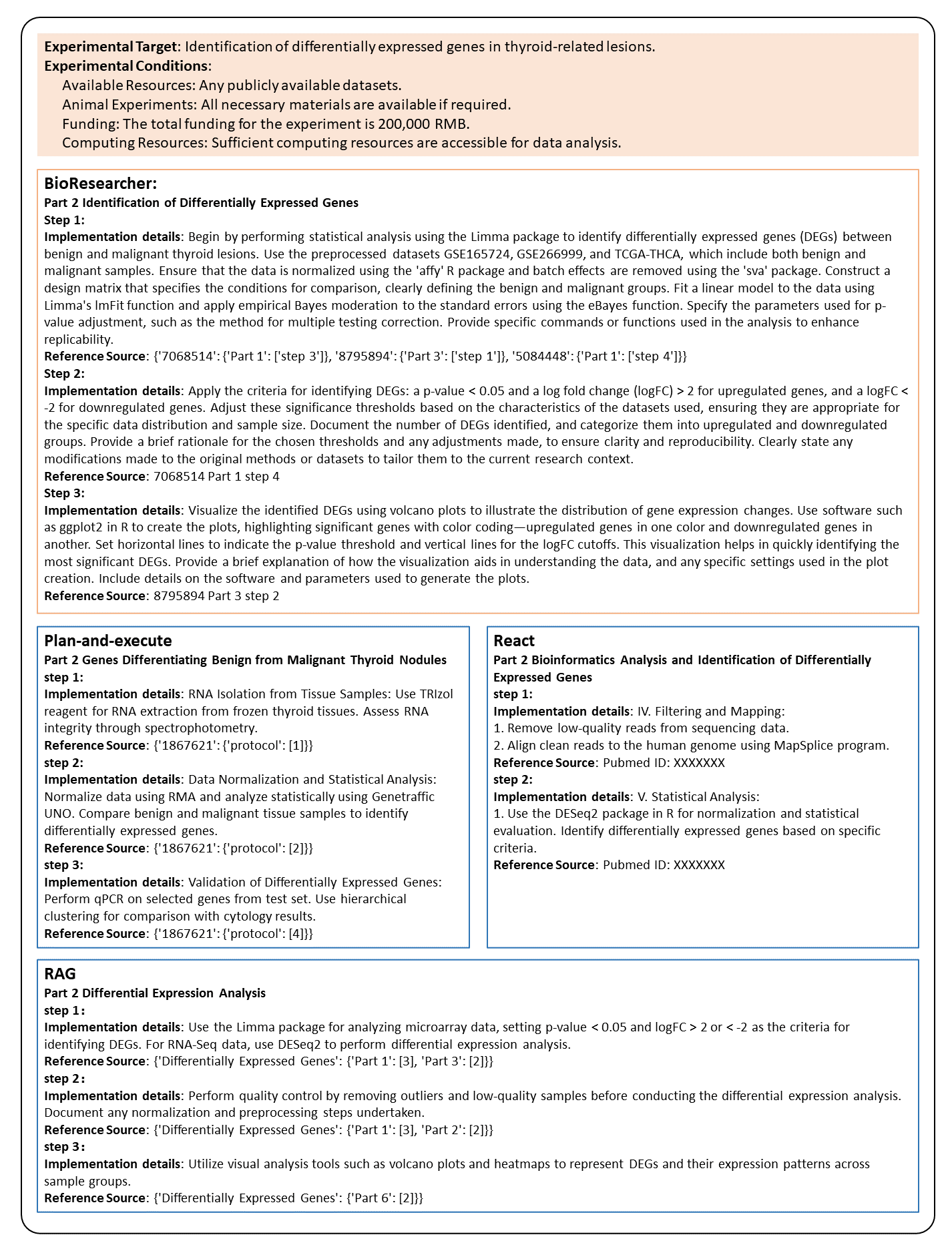}
  \caption{A comparison of \system, Plan-and-Execute, React, and RAG in generating protocols.}
  \label{fig:bad_case}
\end{figure}

Figure~\ref{fig:bad_case} shows a comparison of a part of the experimental protocols generated by four different systems: \system, Plan-and-execute, React, and RAG. Notably, the protocol produced by \system~is both detailed and accurate. In contrast, the protocols generated by the other systems lack essential details, rendering them impractical for implementation.

Table~\ref{tab: protocol_all_models} presents the evaluation results of the four assessment models on the protocols described in Section 5.2. As shown in the table, our method consistently outperforms the baseline systems across all four evaluation models.
\definecolor{darkgreen}{rgb}{0.0, 0.5, 0.0}
\begin{table}[h!]
    \centering
    \caption{Quality of experimental protocols generated by various systems evaluated by different models.}
    \label{tab: protocol_all_models}
    \resizebox{\textwidth}{!}{
    \begin{tabular}{llcccccc|cc}
    \toprule
        Evaluator & Method & Completeness & Detail & Correctness & Logical Soundness & Structure & Overall & $l_{steps}$ & $n_{total~steps}$\\ 
        \midrule
\multirow{4}{*}{GPT-4o} 
& RAG & 0.392  & 0.765  & 0.470  & 0.991  & \textbf{0.952}  & 3.570 & 1.286 & 9.167  \\
& ReAct & 0.361 & 0.617 & 0.487  & 0.972 & 0.950 & 3.387 & 1.237  & 5.792 \\
& Plan and Execute & 0.366 & 0.617 & 0.487  & \textbf{1.000} & 0.935 & 3.405 & 1.194 & 6.375 \\
& \system~&  \textbf{0.612}$_{{\uparrow\textbf{0.220}}}$ & \textbf{0.902}$_{{\uparrow\textbf{0.137}}}$ & \textbf{0.944}$_{{\uparrow\textbf{0.457}}}$ & 0.987$_{\downarrow0.013}$ & 0.910$_{\downarrow0.042}$ & \textbf{4.355}$_{{\uparrow\textbf{0.785}}}$ & 7.327 & 33.958 \\ 
        \midrule
\multirow{4}{*}{o3-mini-3-31} 
& RAG & 0.376  & 0.731  & 0.488  & 0.960  & \textbf{0.967}  & 3.522 & 1.286 & 9.167  \\
& ReAct & 0.329 & 0.608 & 0.479  & 0.958 & 0.956 & 3.330 & 1.237  & 5.792 \\
& Plan and Execute & 0.375 & 0.621 & 0.479  & \textbf{0.943} & 0.960 & 3.378 & 1.194 & 6.375 \\
& \system~& \textbf{0.509}$_{{\uparrow\textbf{0.133}}}$ & \textbf{0.981}$_{{\uparrow\textbf{0.250}}}$ & \textbf{0.823}$_{{\uparrow\textbf{0.435}}}$ & 0.944$_{\downarrow0.017}$ & 0.952$_{\downarrow0.015}$ & \textbf{4.309}$_{{\uparrow\textbf{0.931}}}$ & 7.327 & 33.958 \\ 
        \midrule
\multirow{4}{*}{gemini-2.0-flash} 
& RAG & 0.397  & 0.567  & 0.488  & 0.982  & \textbf{0.820}  & 3.254 & 1.286 & 9.167  \\
& ReAct & 0.329 & 0.455 & 0.490  & 0.966 & 0.795 & 3.036 & 1.237  & 5.792 \\
& Plan and Execute & 0.341 & 0.475 & 0.488  & \textbf{0.971} & 0.827 & 3.102 & 1.194 & 6.375 \\
& \system~& \textbf{0.758}$_{{\uparrow\textbf{0.361}}}$ & \textbf{0.754}$_{{\uparrow\textbf{0.188}}}$ & \textbf{0.872}$_{{\uparrow\textbf{0.382}}}$ & 0.943$_{\downarrow0.039}$ & 0.792$_{\downarrow0.035}$ & \textbf{4.118}$_{{\uparrow\textbf{0.864}}}$ & 7.327 & 33.958 \\
        \midrule
\multirow{4}{*}{deepseek-V3} 
& RAG & 0.454  & 0.685  & 0.486  & 0.960  & \textbf{0.892}  & 3.476 & 1.286 & 9.167  \\
& ReAct & 0.435 & 0.629 & 0.478  & 0.957 & 0.885 & 3.385 & 1.237  & 5.792 \\
& Plan and Execute & 0.437 & 0.633 & 0.476  & \textbf{0.945} & 0.879 & 3.370 & 1.194 & 6.375 \\
& \system~& \textbf{0.758}$_{{\uparrow\textbf{0.305}}}$ & \textbf{0.933}$_{{\uparrow\textbf{0.248}}}$ & \textbf{0.843}$_{{\uparrow\textbf{0.357}}}$ & 0.940$_{\downarrow0.017}$ & 0.910$_{\downarrow0.019}$ & \textbf{4.380}$_{{\uparrow\textbf{0.909}}}$ & 7.327 & 33.958 \\
    \midrule
  \multirow{4}{*}{Average} &  RAG & 0.405  & 0.687  & 0.483  & \textbf{0.973}  & \textbf{0.908}  &  3.456 &1.286 & 9.167  \\
&ReAct & 0.364 & 0.577 & 0.484  & 0.963 & 0.897 & 3.285 & 1.237  &5.792 \\
&Plan and Execute & 0.380 & 0.587 & 0.483  & 0.965 & 0.900 & 3.314 & 1.194 &6.375 \\ 
&\system~&  $\textbf{0.659}_{{\uparrow\textbf{0.254}}}$ & $\textbf{0.893}_{{\uparrow\textbf{0.206}}}$ & $\textbf{0.895}_{{\uparrow\textbf{0.411}}}$ & $0.953_{\downarrow0.020}$& $ 0.891_{\downarrow0.017}$ & $\textbf{4.292}_{{\uparrow\textbf{0.836}}}$ & 7.327 & 33.958\\\bottomrule
    \end{tabular}
    }
\end{table}

\end{appendix}


\begin{thebibliography}{99}


\bibitem{1} National Research Council (US) Committee for Monitoring the Nation's Changing Needs for Biomedical, Behavioral, and Clinical Personnel. Advancing the Nation’s Health Needs: NIH Research Training Programs. Washington (DC): National Academies Press (US), 2005.
\bibitem{2} Huang Y, Liu Y, Pandey N K, et al. Iron oxide nanozymes stabilize stannous fluoride for targeted biofilm killing and synergistic oral disease prevention. Nature Communications, 2023, 14: 6087.
\bibitem{3} Wang P, Song M, Eliassen A H, et al. Optimal dietary patterns for prevention of chronic disease. Nature Medicine, 2023, 29: 719–728.
\bibitem{4} Agarwal A, Mehta P M, Jacobson T, et al. Fixed-dose combination therapy for the prevention of atherosclerotic cardiovascular disease. Nature Medicine, 2024, 30: 1199–1209.
\bibitem{5} Therriault J, Janelidze S, Benedet A L, et al. Diagnosis of alzheimer’s disease using plasma biomarkers adjusted to clinical probability. Nature Aging, 2024, 4: 1529–1537.
\bibitem{6} Zheng J, Sun Q, Zhang M, et al. Noninvasive, microbiome-based diagnosis of inflammatory bowel disease. Nature Medicine, 2024, 30: 3555-3567.
\bibitem{7} Sepahi N, Samsami S, Mansoori Y, et al. Development of a novel colorimetric assay for the rapid diagnosis of coronavirus disease 2019 from nasopharyngeal samples. Scientific Reports, 2024, 14: 12125.
\bibitem{8} Chang V K, Imperial M Z, Phillips P P, et al. Risk-stratified treatment for drug-susceptible pulmonary tuberculosis. Nature Communications, 2024, 15: 9400.
\bibitem{9} Apperloo E M, Gorriz J L, Soler M J, et al. Semaglutide in patients with overweight or obesity and chronic kidney disease without diabetes: a randomized double-blind placebo-controlled clinical trial. Nature Medicine, 2025, 31: 278-285.
\bibitem{10} Douvaras P, Buenaventura D F, Sun B, et al. Ready-to-use ipsc-derived microglia progenitors for the treatment of cns disease in mouse models of neuropathic mucopolysaccharidoses. Nature Communications, 2024, 15: 8132.
\bibitem{11} Gupta R, Srivastava D, Sahu M, et al. Artificial intelligence to deep learning: machine intelligence approach for drug discovery. Molecular Diversity, 2021, 25: 1315–1360.
\bibitem{12} Zhu J, Wang J X, Wang X, et al. Prediction of drug efficacy from transcriptional profiles with deep learning. Nature Biotechnology, 2021, 39: 1444–1452.
\bibitem{13} Mak K K, Wong Y H, Pichika M R. Artificial Intelligence in Drug Discovery and Development. In: Hock F J, Pugsley M K, eds. Drug Discovery and Evaluation: Safety and Pharmacokinetic Assays. Cham: Springer, 2024. 1461–1498.
\bibitem{14} Bizzo B C, Almeida R R, Alkasab T K. Artificial intelligence enabling radiology reporting. Radiologic Clinics, 2021, 59: 1045-1052.
\bibitem{15} European Society of Radiology (ESR). What the radiologist should know about artificial intelligence–an ESR white paper. Insights into Imaging, 2019, 10: 44.
\bibitem{16} Li C T, Zhang Y M, Weng Y, et al. Natural language processing applications for computer-aided diagnosis in oncology. Diagnostics, 2023, 13: 286.
\bibitem{17} Nordin N, Zainol Z, Noor M H M, et al. An explainable predictive model for suicide attempt risk using an ensemble learning and shapley additive explanations (shap) approach. Asian Journal of Psychiatry, 2023, 79: 103316.
\bibitem{18} Malibari A A. An efficient iot-artificial intelligence-based disease prediction using lightweight cnn in healthcare system. Measurement Sensors, 2023, 26: 100695.
\bibitem{19} Bom H S H. Exploring the opportunities and challenges of chatgpt in academic writing: a roundtable discussion. Nuclear Medicine and Molecular Imaging, 2023, 57: 165–167.
\bibitem{20} Lingard L. Writing with chatgpt: An illustration of its capacity, limitations \& implications for academic writers. Perspectives on Medical Education, 2023, 12: 261.
\bibitem{21} Sami A M, Rasheed Z, Kemell K K, et al. System for systematic literature review using multiple ai agents: Concept and an empirical evaluation. arXiv Preprint, 2024, arXiv:2403.08399.
\bibitem{22} Gao S H, Fang A, Huang Y P, et al. Empowering biomedical discovery with ai agents. Cell, 2024, 187: 6125–6151.
\bibitem{23} Chen W J, Cheng J, Cai Y Q, et al. The pyroptosis-related signature predicts prognosis and influences the tumor immune microenvironment in dedifferentiated liposarcoma. Open Medicine, 2024, 19: 20230886.
\bibitem{24} Lin C Y. Rouge: A package for automatic evaluation of summaries. Text Summarization Branches Out, 2004: 74-81.
\bibitem{25} Papineni K, Roukos S, Ward T, et al. Bleu: a method for automatic evaluation of machine translation. In: Proceedings of the 40th Annual Meeting of the Association for Computational Linguistics, 2002: 311-318.
\bibitem{26} Gao Y F, Xiong Y, Gao X Y, et al. Retrieval-augmented generation for large language models: A survey. arXiv Preprint, 2023, arXiv:2312.10997, 2.
\bibitem{27} Wang L, Xu W Y, Lan Y H, et al. Plan-and-solve prompting: Improving zero-shot chain-of-thought reasoning by large language models. In: Rogers A, Boyd-Graber J, Okazaki N, eds. Proceedings of the 61st Annual Meeting of the Association for Computational Linguistics, Toronto, Canada, 2023, 1: 2609–2634.
\bibitem{28} Lu C, Lu C, Lange R T, et al. The ai scientist: Towards fully automated open-ended scientific discovery. arXiv Preprint, 2024, arXiv:2408.06292.
\bibitem{29} Weng Y, Zhu M, Bao G, et al. Cycleresearcher: Improving automated research via automated review. In: The Thirteenth International Conference on Learning Representations, Singapore, 2025.
\bibitem{30} Baek J, Jauhar S K, Cucerzan S, et al. Researchagent: Iterative research idea generation over scientific literature with large language models. In: Chiruzzo L, Ritter A, Wang L, eds. Proceedings of the 2025 Conference of the Nations of the Americas Chapter of the Association for Computational Linguistics: Human Language Technologies, Albuquerque, New Mexico, USA, 2025, 1: 6709–6738.

\bibitem{31} Jin Q, Leaman R, Lu Z. Pubmed and beyond: biomedical literature search in the age of artificial intelligence. EBioMedicine, 2024, 100: 104988.
\bibitem{32} Wang X, Li J, Wu X, et al. Huatuo-26M, a Large-scale Chinese Medical QA Dataset. In: Chiruzzo L, Ritter A, Wang L, eds. Findings of the Association for Computational Linguistics: Annual Conference of the Nations of the Americas Chapter of the Association for Computational Linguistics, Albuquerque, New Mexico, USA, 2025. 3828–3848.

\bibitem{33} Sudarshan M, Shih S, Yee E, et al. Agentic llm workflows for generating patient-friendly medical reports. arXiv Preprint, 2024, arXiv:2408.01112.
\bibitem{34} Devlin J, Chang M W, Lee K, et al. Bert: Pre-training of deep bidirectional transformers for language understanding. In: Ammar W, Louis A, Mostafazadeh N, eds. Proceedings of the 2019 Conference of the North American Chapter of the Association for Computational Linguistics: Human Language Technologies, Minneapolis, Minnesota, 2019, 1: 4171-4186.

\bibitem{35} Singhal K, Azizi S, Tu T, et al. Large language models encode clinical knowledge. Nature, 2023, 620: 172–180.
\bibitem{36} Zhang H, Chen J, Jiang F, et al. HuatuoGPT, Towards Taming Language Model to Be a Doctor. In: Bouamor H, Pino J, Bali K, eds. Proceedings of the 2023 Conference on Empirical Methods in Natural Language Processing, Singapore, 2023, 10859–10885.

\bibitem{37} Chen Y R, Wang Z Y, Xing X F, et al. Bianque: Balancing the questioning and suggestion ability of health llms with multi-turn health conversations polished by chatgpt. arXiv Preprint, 2023, arXiv:2310.15896.
\bibitem{38} Bolton E, Venigalla A, Yasunaga M, et al. Biomedlm: A 2.7 b parameter language model trained on biomedical text. arXiv Preprint, 2024, arXiv:2403.18421.
\bibitem{39} Li Q, Yang X Y, Wang H W, et al. From beginner to expert: Modeling medical knowledge into general llms. arXiv Preprint, 2023, arXiv:2312.01040.
\bibitem{40} Hu E J, Shen Y, Wallis P, et al. Lora: Low-rank adaptation of large language models. In: The Tenth International Conference on Learning Representations, Virtual Event, 2022, 1: 3.
\bibitem{41} Ouyang L, Wu J, Jiang X, et al. Training language models to follow instructions with human feedback. In: Koyejo S, Mohamed S, Agarwal A, Belgrave D, Cho K, Oh A, eds. Advances in Neural Information Processing Systems, 2022, 35: 27730–27744.

\bibitem{42} White J, Fu Q, Hays S, et al. A prompt pattern catalog to enhance prompt engineering with chatgpt. In: Yoder J, Gabriel R, Vranić V, Brown K, eds. Proceedings of the 30th Conference on Pattern Languages of Programs, Monticello IL, USA, 2023, 5, 1–31.

\bibitem{43} Maharjan J, Garikipati A, Singh N P, et al. Openmedlm: prompt engineering can out-perform fine-tuning in medical question-answering with open-source large language models. Scientific Reports, 2024, 14: 14156.
\bibitem{44} Nachane S S, Gramopadhye O, Chanda P, et al. Few shot chain-of-thought driven reasoning to prompt LLMs for open-ended medical question answering. In: Al-Onaizan Y, Bansal M, Chen Y, eds. Findings of the Association for Computational Linguistics: Conference on Empirical Methods in Natural Language Processing, Miami, Florida, USA, 2024, 542–573.

\bibitem{45} Wang J, Shi E, Yu S, et al. Prompt engineering for healthcare: Methodologies and applications. arXiv Preprint, 2023, arXiv:2304.14670.
\bibitem{46} Boiko D A, MacKnight R, Kline B, et al. Autonomous chemical research with large language models. Nature, 2023, 624: 570–578.
\bibitem{47} Nori H, Lee Y T, Zhang S, et al. Can generalist foundation models outcompete special-purpose tuning? case study in medicine. arXiv Preprint, 2023, arXiv:2311.16452.
\bibitem{48} Park J S, O'Brien J, Cai C J, et al. Generative agents: Interactive simulacra of human behavior. In: Follmer S, Han J, Steimle J, Riche N H, eds. Proceedings of the 36th annual ACM Symposium on User Interface Software and Technology, San Francisco, CA, USA, 2023: 1-22.

\bibitem{49} Wang G, Xie Y, Jiang Y, et al. Voyager: An open-ended embodied agent with large language models. Transactions on Machine Learning Research, 2024.
\bibitem{50} Fernando C, Banarse D, Michalewski H, et al. Promptbreeder: Self-referential self-improvement via prompt evolution. In: Salakhutdinov R, Kolter Z, Heller K, Weller A, Oliver N, Scarlett J, Berkenkamp F, eds. Proceedings of the 41st International Conference on Machine Learning, Vienna, Austria, 2024, 235: 13481–13544.

\bibitem{51} Yang C, Wang X, Lu Y, et al. Large language models as optimizers. In: The Twelfth International Conference on Learning Representations, Vienna, Austria, 2024.
\bibitem{52} Tang X R, Zou A N, Zhang Z S, et al. Medagents: Large language models as collaborators for zero-shot medical reasoning. In: Ku L, Martins A, Srikumar V, eds. Findings of the Association for Computational Linguistics, Bangkok, Thailand, 2024, 599–621.

\bibitem{53} Fan Z H, Wei L, Tang J L, et al. AI Hospital: Benchmarking Large Language Models in a Multi-agent Medical Interaction Simulator. In: Rambow O, Wanner L, Apidianaki M, Al-Khalifa H, Eugenio B D, Schockaert S, eds. Proceedings of the 31st International Conference on Computational Linguistics, Abu Dhabi, UAE, 2025, 10183–10213.

\bibitem{54} Li J K, Wang S Y, Zhang M, et al. Agent hospital: A simulacrum of hospital with evolvable medical agents. arXiv Preprint, 2024, arXiv:2405.02957.
\bibitem{55} LI G. Ai4r: The fifth scientific research paradigm. Bulletin of Chinese Academy of Sciences, 2024, 39: 1–9.
\bibitem{56} Lim Y, Tamayo-Orrego L, Schmid E, et al. In silico protein interaction screening uncovers donson’s role in replication initiation. Science, 2023, 381: eadi3448.
\bibitem{57} Zhou J, Zhang B, Chen X, et al. Automated bioinformatics analysis via autoba. arXiv Preprint, 2023, arXiv:2309.03242.
\bibitem{58} M. Bran A, Cox S, Schilter O, et al. Augmenting large language models with chemistry tools. Nature Machine Intelligence, 2024, 6: 525-535.
\bibitem{59} Huang K, Qu Y, Cousins H, et al. Crispr-gpt: An llm agent for automated design of gene-editing experiments. arXiv Preprint, 2024, arXiv:2404.18021.
\bibitem{60} Tiukova I A, Brunnsaker D, Bjurstr¨om E Y, et al. Genesis: Towards the automation of systems biology research. arXiv Preprint, 2024, abs/2408.10689.
\bibitem{61} Su H, Chen R, Tang S, et al. Two heads are better than one: A multi-agent system has the potential to improve scientific idea generation. arXiv Preprint, 2024, arXiv:2410.09403.
\bibitem{62} Yang X, Chen H, Feng W, et al. Collaborative evolving strategy for automatic data-centric development. arXiv Preprint, 2024, arXiv:2407.18690.
\bibitem{63} Clark J. Systematic Reviewing. In: Doi S, Williams G, eds. Methods of Clinical Epidemiology. Berlin, Heidelberg: Springer, 2013. 187–211.
\bibitem{64} Scells H, Zuccon G, Koopman B, et al. Automatic boolean query formulation for systematic review literature search. In: Huang Y, King I, Liu T, Steen M V, eds. Proceedings of the Web Conference, Taipei Taiwan, 2020, 1071-1081.

\bibitem{65} Bai Y S, Lv X, Zhang J J, et al. LongBench: A Bilingual, Multitask Benchmark for Long Context Understanding. In: Ku L, Martins A, Srikumar V, eds. Proceedings of the 62nd Annual Meeting of the Association for Computational Linguistics, Bangkok, Thailand, 2024, 1: 3119–3137.

\bibitem{66} Wei J, Wang X, Schuurmans D, et al. Chain-of-thought prompting elicits reasoning in large language models. In: Koyejo S, Mohamed S, Agarwal A, Belgrave D, Cho K, Oh A, eds. Advances in Neural Information Processing Systems, 2022, 35: 24824-24837.

\bibitem{67} Begley C G, Ioannidis J P. Reproducibility in science: improving the standard for basic and preclinical research. Circulation Research, 2015, 116: 116–126.
\bibitem{68} Larijani B, Tayanloo-Beik A, Payab M, et al. Design of experimental studies in biomedical sciences. Biomedical Product Development: Bench to Bedside, 2020, 37-47.
\bibitem{69} Yao S, Zhao J, Yu D, et al. React: Synergizing reasoning and acting in language models. In: The Eleventh International Conference on Learning Representations, Kigali, Rwanda, 2023.

\bibitem{70} Arnold C. Cloud labs: where robots do the research. Nature, 2022, 606: 612-613.


\bibitem{71} Wu Q Y, Bansal G, Zhang J Y, et al. Autogen: Enabling next-gen LLM applications via multi-agent conversation framework. arXiv preprint, 2023, abs/2308.08155.

\end{thebibliography}
\end{document}